\documentclass[11pt,a4paper]{article}
\pdfoutput=1

\usepackage[%
bibstyle=numeric,%
style=numeric-comp,%
sorting=none,%
sortcites=true,%
maxnames=4,%
abbreviate=true,%
backend=bibtex,%
]{biblatex}

\bibliography{GaugeBetheRef}
\usepackage[]{Packages}

\usepackage{Definitions}

\allowdisplaybreaks[1]

\preprint{\texttt{CERN-PH-TH/2012-090}\\\texttt{IPMU12-0069}}

\newcommand{\OfficialTitle}{The Omega Deformation\\ from String and M--Theory}

\title{\vspace{2cm}
  {\huge   \textbf{\sffamily\OfficialTitle}}
}

\hypersetup{pdfauthor={Simeon Hellerman and Domenico Orlando and Susanne Reffert},pdftitle={\OfficialTitle}}

\author{
  \begin{minipage}{.8\linewidth}
    \vspace{1cm}
    \begin{center}
      {\small \textbf{Simeon Hellerman}\( {}^\flat \), \textbf{Domenico Orlando}\( {}^\# \) and \textbf{Susanne Reffert}\( {}^\# \) }
    \end{center}
    \vspace{1cm}
    \begin{minipage}{\linewidth}\centering
      {\itshape \footnotesize 
        \begin{itemize}
        \item[\( {}^\flat \)] Kavli Institute for the Physics and Mathematics of the Universe (Kavli IPMU),\\ The University of Tokyo (ToDIAS), \\ Kashiwa, Chiba 277-8568, Japan.
        \item[\( {}^\# \)] Theory Group, Physics Department, \\ Organisation européenne pour la recherche nucléaire (CERN) \\ CH-1211 Geneva 23, Switzerland
        \end{itemize}
      }
    \end{minipage}
  \end{minipage}
}

\date{}

\begin{document}

\setstretch{1.1}

\numberwithin{equation}{section}

\begin{titlepage}

  \maketitle

  \thispagestyle{empty}

  \vfill
  \abstract{{We present a string theory construction of Omega-deformed four-dimensional gauge theories with generic values of $\epsilon_1$ and $\epsilon_2$.  Our solution gives an explicit description of the geometry in the core of Nekrasov and Witten's realization of the instanton partition function, far from the asymptotic region of their background.  This construction lifts naturally to M--theory and corresponds to an M5--brane wrapped on a Riemann surface with a selfdual flux. Via a 9--11 flip, we finally reinterpret the Omega deformation in terms of non-commutative geometry.  Our solution generates all modified couplings of the $\Omega$--deformed gauge theory,  and also yields a geometric origin for the quantum spectral curve of the associated quantum integrable system.} }
\vfill

\end{titlepage}

\section{Introduction}

The $\Omega$--deformation~\cite{Moore:1997dj, Lossev:1997bz} has been a topic of intense research in the past years.  While it first appeared in the context of instanton sums~\cite{Nekrasov:2002qd}, it has also received attention from the topological string community~\cite{Hollowood:2003cv, Iqbal:2007ii, Antoniadis:2010iq, Krefl:2010fm,Huang:2010kf,Bonelli:2011na,Aganagic:2011mi} and more recently in connection with integrable systems via the gauge/Bethe correspondence~\cite{Nekrasov:2009ui,Nekrasov:2009rc,Orlando:2010uu,Orlando:2010aj}.

In this paper, we will present a generally applicable formulation of the $\Omega$--de\-for\-ma\-tion from string theory via a brane construction in the so-called \emph{fluxtrap} background, as first presented in~\cite{Hellerman:2011mv} and generalized further in~\cite{Reffert:2011dp, Orlando:2011nc}. We will demonstrate the versatility of the fluxtrap approach by connecting to topics discussed in the recent literature, such as the (refined) topological string and the so-called \ac{ns} limit appearing in the context of the four-dimensional gauge/Bethe correspondence~\cite{Nekrasov:2009rc}. As we will show, also the M--theory lift of $\Omega$--deformed theories can be performed in our fluxtrap setup with relative ease.

As an explicit application,  we show how the non-commutative spectral curve of a quantum integrable system emerges geometrically from the \ac{ns}~gauge theory via a 9--11 flip. A similar non-commutativity also appears in the related limit of the $\Omega$--deformation corresponding to the topological string~\cite{Aganagic:2003qj, Dijkgraaf:2007sw}.  In our case the geometric
interpretation of the quantization is more direct, with the symplectic form on the
curve realized as the pullback of M-theory four-form flux to an M5-brane.

\bigskip

The fluxtrap background~\cite{Hellerman:2011mv} is the T--dual of a \emph{fluxbrane} or \emph{Melvin} background. This is an integrable string theory, for it is a free quotient that can be studied with the methods of~\cite{Tseytlin:1994ei,Tseytlin:1995zv,Russo:2001na,Gutperle:2001mb,Takayanagi:2001jj, Hellerman:2006tx}. Depending on the configuration of
branes which is placed in it, the fluxtrap can give rise to both \emph{twisted masses} in a two-dimensional gauge theory~\cite{Hellerman:2011mv}, or an $\Omega$--deformed four-dimensional gauge theory (which is effectively dimensionally-reduced by the deformation). The number of $\epsilon$--parameters by which the gauge theory is deformed is restricted by the number of available isometries in the undeformed metric. Supersymmetry moreover imposes a relation on the $\epsilon_i$. The case $\epsilon_1=-\epsilon_2$ related to the topological string and the case $\epsilon_1=0$ corresponding to the \ac{ns}~limit are both special limits of the general construction discussed here.

\bigskip

The plan of the paper is as follows. In Section~\ref{sec:fluxtrap}, we introduce the
construction of the Nekrasov $\Omega$--deformation of an $\mathcal{N}=2$ gauge
theory as the Melvin compactification of a $(p,q)$ fivebrane web, whose
T--dual description is the flux-trap of \cite{Hellerman:2011mv}; 
in this section we also review the fluxtrap construction in its most general formulation on a generic Ricci-flat space. In Section~\ref{sec:omega-def}, we introduce the brane setups which gives rise to a variety of $\Omega$--deformed gauge theories.  In Section~\ref{sec:m-theory-lift}, the M--theory lift of our brane setup in the fluxtrap background is discussed. In Section~\ref{sec:9-11}, we perform a 9--11 flip to realize the non-commutative spectral curve of the quantum integrable systems associated to topological strings and gauge theories in the \ac{ns} limit. Conclusions and outlook are given in Section~\ref{sec:conc}.

In Appendix~\ref{sec:supersymm-conv}, some supersymmetry conventions which are used throughout the article are collected, while in Appendix~\ref{sec:taub-nut-coordinates} the \ac{tn} space is presented in different coordinate systems which become useful in different parts of this article.
While it is possible to deform by complex $\epsilon$--parameters~\cite{Reffert:2011dp}, we restrict ourselves in this article to $\epsilon\in\mathbb{R}$, partly for ease of exposition and partly because it is the more natural choice under some circumstances. The general set-up for the complex fluxtrap background is however formally appealing and is presented for completeness in Appendix~\ref{sec:compeps}.

\section{Omega-deformation of \acl{sw} theory as a fivebrane web in type IIB Melvin background}\label{sec:fluxtrap}

We wish to study the $\Omega$--deformation of a general 
$\mathcal{N} = 2$ gauge theory in four dimensions that has a realization
in the manner of~\cite{Witten:1997sc}, as the low-energy dynamics of 
a set of \NS5 and \D4 branes of \tIIA string theory,
that can be deformed to a stack of M--fivebranes
wrapped on a Riemann surface in M--theory.  The $\Omega$--deformation of a four-dimensional
theory does not have a simple, universal description in terms of the four-dimensional
dynamics itself, but can generally be understood most simply by lifting to a five-dimensional
supersymmetric theory on a circle, and then re-compactifying on a circle
with certain twisted boundary
conditions~\cite{Nekrasov:2003rj}.  The most direct way to realize the $\Omega$--deformed
four-dimensional gauge theory into string theory, then, is to lift this general prescription to string theory.
This will turn out to be quite straightforward, with the compactification from five to
four dimensions realized as a T--duality of the bulk string theory.

The class of gauge theories in~\cite{Witten:1997sc} generally have lifts to five-dimensional theories that can be realized as the dynamics of a web of \NS5-- and \D5--branes~\cite{Aharony:1997bh,Hanany:1997tb}.  The compactification on the $S^1$ produces a four-dimensional gauge theory realized
on \tIIA \NS5-- and \D4--branes by the usual T--duality rules that turn a longitudinally
compactified \D5--brane into a \D4--brane, and a longitudinally compactified \NS5--brane
into another \NS5--brane.  

While the T--duality of the branes works out in a completely obvious way, the T--duality
of the bulk itself does not, despite the simplicity of the initial background.
The starting point for our solution is a vacuum Einstein metric,
identified by a simultaneous
shift in one direction and a rotation of
some other directions (Melvin background).
In the case where the rotated coordinates are flat Minkowski space,
these solutions are sometimes referred to as ``fluxbranes'' and have long
been studied, starting with~\cite{Melvin:1963qx} and continuing into
the modern era~\cite{Tseytlin:1994ei,Tseytlin:1995zv,Russo:2001na,Gutperle:2001mb,Takayanagi:2001jj}.

We will always be considering identifications of the product of a line with a four-dimensional
base $x\uu{0,1,2,3}$ which will either be flat $\IR\uu 4$ or else a \ac{tn} geometry
with asymptotic radius~$\rTN$.  The first case is a limit of the second as $\rTN \to \infty$, and the distinction between the two will affect nothing relevant to our consideration. 
This is because the
supersymmetrically invariant quantities counted by the Nekrasov partition function are
known~\cite{Gaiotto:2005gf, Dijkgraaf:2006um} not to depend on the radius $\rTN$ as
an independent parameter.

The rotational identification of $x\uu{0,1,2,3}$ is
a rotation in two different planes by angles~$\theta\lll 1$~and~$\theta\lll 2$, and the
translational identification on the real line is a shift by an amount $\wt R$.  The limit
of interest to us is the limit where both $\wt R \to 0 $ and $ \theta\lll{1,2} \to 0$ with~$\theta\lll {1,2}/ \wt R \equiv \epsilon\lll {1,2}$ are held fixed.
The resulting spacetime, thought of as a fibration of $x\uu{0,1,2,3}$ over $\tilde x^9$
is locally trivial, but it is not locally trivial when thought of as a fibration of 
$\tilde x^9$ over $x\uu{0,1,2,3}$.  
In a gravitational theory on this spacetime, the local nontriviality of the fibration manifests
itself as a nonzero electromagnetic field strength of the Kaluza--Klein gauge connection.

In the low-energy effective quantum theory of gravity, the limit $\wt R \to 0$ is a singular
one: quantum effects are not under control when $\wt R$ becomes
smaller than the scale of new states and/or nonrenormalizable interactions.  When the effective theory is embedded in \tIIB string theory, as is the case for us, the scale of
new states and nonrenormalizable interactions is the string length
$\ell\lll s = \sqrt{\alpha^\prime}$,
below which light winding states dominate the spectrum and the controlled description is
in terms of a T--dual theory of branes in \tIIA, rather than \tIIB string theory. This
T--duality turns the Kaluza--Klein electromagnetic flux into curvature~$H$
of the Neveu--Schwarz $B$--field, meaning that a noncommutative 
deformation of the brane dynamics enters the description as an essential part of
the $\Omega$--deformation.  We now give the details of this solution and
its description in various dual frames.

The deformation with general $\e$--parameters rotates the two complex planes near the 
core of the \ac{tn} geometry with arbitrary angles as one traverses the Melvin circle.
Generically, this would by itself break all supersymmetry, so one needs to extend the
deformation by a third rotation of the spacetime that also acts with a nontrivial phase
on spinors.  
Such a transverse rotational isometry always exists in the cases we consider:
For a generic ${\cal N} = 2$ $(p,q)$ fivebrane web preserving
eight supercharges, there are three common transverse directions that 
rotate into one another under an $SO(3)$ transverse isometry.  This 
isometry acts as $SU(2)$ on the Killing spinors preserved by the fivebranes and
is thus an exact $SU(2)$ R--symmetry, not only at low energies but
exact in the full ultraviolet-complete string dynamics.  We shall return to this point
later when we compare our realization of the $\Omega$--deformed gauge
theory with its other well known realization as the topological string \cite{Nekrasov:2002qd,Iqbal:2003ix}.

\paragraph{The background.}

Let us consider a \tIIB background given by a (Euclidean) Ricci-flat metric of the type \( \di s^2 = g_{ij} \di x^i \di x^j + \di (\tilde x^9)^2 \) and a constant dilaton \( \Phi_0 \). The direction \( \tilde x^9 = \wt R \tilde u \) describes a circle of radius \( \wt R \) and the metric \( g \) has \( N \le 4 \) (non-compact) rotational isometries generated by \( \del_{\theta_k} \). The Melvin identifications on this background are
\begin{align}\label{eq:melv}
  \begin{cases}
    \tilde u \sim \tilde u + 2 \pi n_u \, , \\
    \theta_k \sim \theta_k + 2 \pi \epsilon_k \wt R n_u\,,
  \end{cases} && n_u \in \setZ
\end{align}
together with the standard identifications \( \theta_k \sim \theta_k + 2 \pi n_k \) for the angular variables. It is convenient to pass to a set of disentangled variables
\begin{equation}
  \phi_k = \theta_k - \epsilon_k \wt R  \tilde u \, ,
\end{equation}
which are \( 2 \pi \)-periodic. The change of variables modifies the boundary conditions from
\begin{equation}
\label{eq:Melvin-preduality}
  \textstyle{ ( \tilde u, \theta_k ) \sim ( \tilde u, \theta_k ) + 2 \pi n_u \left( 1, \epsilon_k \wt R \right) +  2 \pi n_k (0,1)}
\end{equation}
to
\begin{equation}
  ( \tilde u, \phi_k ) \sim (\tilde u, \phi_k) + 2 \pi n_u \left( 1, 0 \right) + 2 \pi n_k (0, 1)\,.
\end{equation}

T--duality in \( \tilde u \) leads to a background with a non-trivial dilaton and a \( B \)--field where all the relevant degrees of freedom are local fields rather than winding strings. We call this the \emph{fluxtrap background on \( g \)}: 
\begin{subequations}
  \label{eq:general-fluxtrap}
  \begin{align}
    \di s^2 &= g_{ij} \di x^i\di x^j  + \frac{(\di x^9)^2 - \epsilon^2 U_i U_j \di x^i \di x^j}{1 + \epsilon^2 \|U\|^2}  \, , \\
    B&= \epsilon \frac{U_i\di  x^i\wedge \di x^9}{1 + \epsilon^2 \|U\|^2} \, ,\\
    \eu^{-\Phi} &= \frac{\sqrt{\alpha'} \eu^{-\Phi_0}}{R}\sqrt{1 + \epsilon^2 \|U\|^2} \, ,
  \end{align}
\end{subequations}
where \( x^9 \) is a circle of radius \( R = \alpha'/ \wt R \),
\begin{equation}
  U^i\del_i = \sum_k \frac{\epsilon_k}{\epsilon} \del_{\phi_k}\,,  
\end{equation}
$U_i = g_{ij}U^j$, and the norm is taken in the initial metric:
\begin{equation}
  \|U\|^2 \equiv U_i U^i \equiv U^i g_{ij} U^j \, .  
\end{equation}
In the limit \( \epsilon \to 0 \) the T--duality is performed on the \( \tilde u \) circle alone and the background remains undeformed.

The Killing vector \( U^i \del_i \) generates the rotational isometries for both the initial and the T--dual metrics. In presence of the fluxtrap the isometry is always bounded, by which we mean that the generating vector field has bounded norm\footnote{In equivariant cohomology, the norm \( \epsilon^2 \|U\|^2 \) is related to the moment map for the rotation generated by~\( U \).}:
\begin{equation}
  \| U \|^2_{\text{trap}} = \frac{\|U\|^2}{1 + \epsilon^2 \|U\|^2} < \frac{1}{\epsilon^2}  \, .
\end{equation}
In this sense \( \epsilon \) acts as a regulator for the non-bounded rotational isometry.

If we set $\Phi_0=\log \wt R/\sqrt{\alpha'}$, which we will do for the rest of this article, we find that the prefactor in the dilaton $\sqrt{\alpha'} \eu^{-\Phi_0} / R = 1$. This will result in the right normalization for the four-dimensional gauge theories in Section~\ref{sec:omega-def}.

\paragraph{Supersymmetry.}

The Melvin identifications on non-bounded isometries break in general all the supersymmetries of the initial Ricci-flat metric. Some of them can be preserved by imposing conditions on the parameters \( \epsilon_k \)~\cite{Russo:2001na}.

In an appropriate coordinate system one can write the Killing spinors \( \etaB \) for the metric \( g_{ij} \di x^i \di x^j + ( \di \tilde x^9)^2 \) in a form that isolates the dependence on the coordinates $\theta_k$ and $\tilde u$:
\begin{equation}
  \etaB = \left( \Id + \Gamma_{11} \right) \prod_{k=1}^N \exp [ \frac{\theta_k}{2} \Gamma_{\rho_k \theta_k} ] \left( \eta_0 + \im \eta_1  \right) \, ,  
\end{equation}
where \( \rho_k, \theta_k \) are the cylindrical coordinates in the plane of the rotation generated by \( \del_{\theta_k} \) and \( \eta_0 + \im \eta_1 \) is a spinor which does not depend on either \( \tilde u \) or \( \theta_k \). The Killing spinor is invariant under \( \theta_k \to \theta_k + 2 \pi n_k \), but not under the Melvin identifications in Eq.~\eqref{eq:Melvin-preduality}. To isolate the source of the problem we pass to the disentangled coordinates \( \phi_k \),
\begin{equation}
  \etaB = \prod_{k=1}^N \exp [ \frac{\phi_k}{2} \Gamma_{\rho_k \theta_k} ] \exp [\frac{\wt R \tilde u}{2} \epsilon_k \Gamma_{\rho_k \theta_k} ] \etaw \, ,
\end{equation}
where \( \etaw = \left( \Id + \Gamma_{11} \right) \left(\eta_0 + \im \eta_1 \right)\). 
While the first term is invariant under \( \phi_k \to \phi_k + 2 \pi n_k \), the second one is not invariant under \( \tilde u \to \tilde u + 2 \pi n_u \). In order to respect the boundary conditions we need to impose
\begin{equation}
  \sum_{k=1}^N \epsilon_k \Gamma_{\rho_k \theta_k} \etaw = 0 \, .  
\end{equation}
This is in general not a projector and all supersymmetries are broken.

Consider now the case \( N > 1 \) and impose the condition
\begin{equation}
  \sum_{k=1}^N s_k \epsilon_k = 0 \, ,  
\end{equation}
where the \( s_k \) are signs. The boundary conditions become
\begin{equation}
  \sum_{k=1}^N \epsilon_k \Gamma_{\rho_k \theta_k} \etaw = \sum_{k=1}^{N-1} \epsilon_k \left( \Gamma_{\rho_k \theta_k} -\frac{s_k}{s_N} \Gamma_{\rho_N \theta_N}  \right) \etaw = 0 \, .  
\end{equation}
What we find is a generic linear combination of \( N - 1 \) commuting projectors. It is annihilated by the product of all the corresponding orthogonal projectors
\begin{equation}
  \proj{flux} = \prod_{k=1}^{N-1} \left( \Gamma_{\rho_k \theta_k} + \frac{s_k}{s_N} \Gamma_{\rho_N \theta_N}  \right) ,
\end{equation}
so that the boundary conditions are satisfied by the Killing spinor
\begin{equation}
  \etaB = \left( \Id + \Gamma_{11} \right) \prod_{k=1}^N \exp [ \frac{\phi_k}{2} \Gamma_{\rho_k \theta_k} ] \proj{flux} \etaw \, .
\end{equation}
Depending on \( \etaw \), the projector \( \proj{flux} \) can either break all supersymmetries or preserve some of them. In the latter case, at least \( 1/2^{N-1} \) of the original ones are preserved.

Since all dependence on \( \tilde u \) has disappeared from the expression, T--duality maps the Killing spinors \( \etaB \) into local \tIIA Killing spinors \( \etaA \). Using an appropriate vielbein for the T--dual metric (see Appendix~\ref{sec:supersymm-conv}) they take the form \( \etaA = \etaA^L + \etaA^R \) with 
\begin{equation}
  \begin{cases}
  \etaA^L = \left(\Id + \Gamma_{11} \right) \displaystyle{\prod_{k=1}^N} \exp [ \frac{\phi_k}{2} \Gamma_{\rho_k \theta_k} ] \proj{flux} \eta_0 \, ,  \\
  \etaA^R = \left( \Id - \Gamma_{11} \right)  \Gamma_{u} \displaystyle{\prod_{k=1}^N} \exp [ \frac{\phi_k}{2} \Gamma_{\rho_k \theta_k} ] \proj{flux} \eta_1 \, ,
  \end{cases}
\end{equation}
where \( \Gamma_u \) is the gamma matrix in the \( u \) direction normalized to unity.
It is possible to write an explicit expression for \( \Gamma_u \), in terms of a rotation on the right-moving spinor that depends on \( \epsilon \). Observe that by construction
\begin{equation}
  \Gamma_u = \frac{\gamma_9 + \epsilon \slashed{U}}{\sqrt{1+ \epsilon^2 \| U \|^2}} \, ,
\end{equation}
where
\begin{equation}
  \slashed U = U^i \EB\indices{^a_i} \gamma_a \, ,  
\end{equation}
and \( \EB \) is the vielbein for the initial Ricci-flat metric. Introducing the angle \( \vartheta \) as
\begin{equation}
  \tan \frac{\vartheta}{2} = \epsilon \| U \| ,
\end{equation}
the gamma matrix becomes
\begin{equation}
  \Gamma_u = \cos \frac{\vartheta}{2} \gamma_9 + \sin \frac{\vartheta}{2} \frac{\slashed{U}}{ \| U \|} = \exp [ \frac{\vartheta}{2} \frac{\slashed{U} \gamma_9}{ \| U \|} ] \gamma_9 \, ,
\end{equation}
where we used the fact that \( \{ \slashed{U}, \gamma_9 \} = 0 \) and \( \slashed{U}^2 = \|U\|^2 \Id \).

\bigskip
  
In conclusion we see that for generic values of \( \epsilon_k \) all supersymmetries are broken. If the sum of the \( \epsilon_k \) is zero, some supersymmetries can be preserved. If this is the case, a minimum of \( 1/2^{N-1} \) of the original supersymmetries are present in the fluxtrap background.

\section{D--branes and Omega--deformations of gauge theories}\label{sec:omega-def}

After having introduced the fluxbrane background in the bulk, we will now study D--brane constructions in this background.
Gauge theories encoding the fluctuations of D--branes placed into a fluxtrap background receive deformations from it. The precise nature of the deformation depends on the type of brane and the way it is placed into the background. \D2--branes suspended between \NS5--branes which are not extended along the planes of rotation for example receive twisted mass deformations~\cite{Hellerman:2011mv}. In this article, we will be concerned with \D4--branes which are extended in the directions of (some of) the rotations. This leads to an $\Omega$--deformation of the resulting gauge theory. Depending on whether one or two planes of rotation lie in the worldvolume of the \D4--branes, we reproduce either the general case $ \epsilon_1 \neq \epsilon_2 $, or special limits such as the $\epsilon_1=-\epsilon_2$ limit which is related to the topological string, or the limit $ \epsilon_1 = 0$, also known as the \acl{ns} limit.

\subsection{General \texorpdfstring{$ \epsilon_1 \neq \epsilon_2 $}{} case}
\label{sec:generaleps}

\paragraph{Closed strings.}

Consider the general construction introduced in the last section for the simplest case of flat space and identifications in three planes. Now \( U \) is the Killing vector corresponding to the rotations in the \( (x^0, x^1) \), \( (x^2, x^3 ) \) and \( (x^4, x^5 ) \) planes,    
\begin{equation}
  \textstyle \epsilon\, U = \epsilon_1 \left( x^0 \del_1 - x^1 \del_0 \right) + \epsilon_2 \left( x^2 \del_3 - x^3 \del_2 \right) + \epsilon_3 \left( x^4 \del_5 - x^5 \del_4 \right) \, ,
\end{equation}
and the fluxtrap background takes the form
\begin{subequations}
  \label{eq:IIA-TwoEpsilonBulk}
  \begin{align}
    \di s^2 &= \di \mathbf{x}_{0\dots8}^2 + \frac{ (\di x^9)^2 - \left( \epsilon_1 \rho_1^2 \di \phi_1 + \epsilon_2 \rho_2^2 \di \phi_2 +  \epsilon_3 \rho_3^2 \di \phi_3 \right)^2}{ 1 + \epsilon_1^2 \rho_1^2 + \epsilon_2^2 \rho_2^2 + \epsilon_3^2 \rho_3^2 } \, , \\
    B &=  \frac{ \left( \epsilon_1 \rho_1^2 \di \phi_1 + \epsilon_2 \rho_2^2 \di \phi_2 + \epsilon_3 \rho_3^2 \di \phi_3  \right) \wedge \di x^9}{ 1 + \epsilon_1^2 \rho_1^2 + \epsilon_2^2 \rho_2^2 + \epsilon_3^2 \rho_3^2} \, , \\
    \eu^{-\Phi} &= \sqrt{ 1 + \epsilon_1^2 \rho_1^2 + \epsilon_2^2 \rho_2^2 + \epsilon_3^2 \rho_3^2} \, .
    \label{eq:IIA-TwoEpsilonDilaton}
  \end{align}
\end{subequations}
In order to preserve some supersymmetry we impose
\begin{equation}
  \epsilon_1 + \epsilon_2 + \epsilon_3 = 0\,,
\end{equation}
and using the general prescription introduced in the previous section it is immediate to see that the background preserves \( 32 / 2^2 = 8 \) supercharges. The Killing spinors are
\begin{equation}
  \label{eq:IIA-TwoEpsilonSpinors}
  \begin{cases}
    \etaA^L = \left(\Id + \Gamma_{11} \right) \exp [ \frac{\phi_1}{2} \gamma_{01} ] \exp [ \frac{\phi_2}{2} \gamma_{23} ] \exp [ \frac{\phi_3}{2} \gamma_{45} ] \left(\gamma_{01} + \gamma_{23} \right) \left( \gamma_{23} + \gamma_{45} \right)\eta_0 \, ,  \\
    \etaA^R = \left(\Id - \Gamma_{11} \right) \Gamma_u \exp [ \frac{\phi_1}{2} \gamma_{01} ] \exp [ \frac{\phi_2}{2} \gamma_{23} ] \exp [ \frac{\phi_3}{2} \gamma_{45} ] \left(\gamma_{01} + \gamma_{23} \right) \left( \gamma_{23} + \gamma_{45} \right)\eta_1 \, ,  \\
  \end{cases}
\end{equation}
where
\begin{equation}
  \Gamma_u =\frac{\gamma_1 \epsilon_1 \rho_1+ \gamma_7 \epsilon_2 \rho_2 + \gamma_5 \epsilon_3 \rho_3 + \gamma_9 }{ \sqrt{ 1 + \epsilon_1^2 \rho_1^2 + \epsilon_2^2 \rho_2^2 + \epsilon_3^2 \rho_3^2}} \, ,
\end{equation}
\( \eta_0 \) and \( \eta_1 \) are constant real spinors (each of the two projectors \( (\gamma_{ij} + \gamma_{kl}) \)  reduces supersymmetry by \( 1/2 \)), and
\begin{align}
  \rho_1 \eu^{\im \phi_1} &= x^0 + \im x^1 \, , & \rho_2 \eu^{\im \phi_2} &= x^2 + \im x^3 \, ,& \rho_3 \eu^{\im \phi_3} &= x^4 + \im x^5 \, . 
\end{align}

\paragraph{Open strings.}

We want to study the embedding of a \D4--brane extended between two \NS5--branes in our background. The \NS5s are extended in the directions \( 012389 \), the \D4 in \( 01236 \), which means that it is finite in the \( x^6 \) direction).  The \ac{dbi} action describes the fluctuations of the \D4 in the directions \( x^8 \) and \( x^9 \), which we collect in a complex field \( v \). 

\bigskip

\begin{table}
  \centering
  \begin{tabular}{lcccccccccc}
    \toprule
    \( x \)   & 0               & 1               & 2               & 3  & 4  & 5  & 6  & 7 & 8  & 9  \\
    \midrule
    fluxbrane & \ep{\epsilon_1} & \ep{\epsilon_2} & \ep{\epsilon_3} & \X & \X & \X & \T               \\
    \NS5      & \X              & \X              & \X              & \X &    &    &    &   & \X & \X \\
    \D4       & \X              & \X              & \X              & \X &    &    & \X &   &    &    \\
    \midrule
    \( \xi \) & 0               & 1               & 2               & 3  &    &    & 4  &   & \ep{v}  \\
    \bottomrule
  \end{tabular}
  \caption{\D4--branes suspended between \NS5s with two independent \( \epsilon \). The crosses \( \X \) indicate directions in which the branes are extended. The circle \( \T \) is the direction of the T--duality.
The effective gauge theory describing the \D4--brane is the \( \Omega \)--deformed four-dimensional gauge system of Nekrasov. Remarkable limits are obtained for \( \epsilon_3 = 0\) (topological strings) and \( \epsilon_1 = 0 \) (the so-called \ac{ns} limit). Note that all directions have the same Euclidean signature.}
  \label{tab:D4-embedding-two-epsilon}
\end{table}

Consider the static embedding defined by
\begin{align}
  f:  \xi^0 &= x^0, &
  \xi^1 &= x^1, &
  \xi^2 &= x^2, &
  \xi^3 &= x^3, &
  \xi^4 &= x^6, &
  v = x^8 + \im x^9 &= v(\xi^0, \xi^1, \xi^2, \xi^3).
\end{align}
The \acl{dbi}~action is given by
\begin{equation}
  S = - \mu_p \int \di ^5 \xi \, \eu^{-\Phi} \sqrt{- \det ( \hat g + \hat B + 2 \pi \alpha' F)}\,  ,
\end{equation}
with $\mu_p = \left( 2 \pi \right)^{-p} \left(\alpha' \right)^{-(p+1)/2}$.
It is convenient to introduce the pullback of the vector field \( U \),
\begin{equation} \label{eq:pullU}
  \textstyle \epsilon\, \hat U = \epsilon f^* U  = \epsilon \hat U^i \del_{\xi^i} = \epsilon_1 \left( \xi^0 \partial_1 - \xi^1 \partial_0 \right) + \epsilon_2 \left( \xi^2 \partial_3 - \xi^3 \partial_2 \right) .
\end{equation}
Expanding the square root of the determinant at second order in the fields, we can write the Lagrangian as
\begin{equation}
  \label{eq:Two-epsilon-gauge}
  \mathscr{L}_{\epsilon_1, \epsilon_2 } = \frac{1}{4 g_4^2}\left[ 1 + F_{ij} F^{ij} + \frac{1}{2} \left( \partial_i \varphi + \im \epsilon \hat U^k F_{ki}  \right) \delta^{ij}  \left( \partial_j \bar \varphi - \im \epsilon \hat U^l F_{lj} \right) 
  - \frac{\epsilon^2}{8} \left( \hat U^i \partial_i ( \varphi + \bar \varphi)  \right)^2 \right],
\end{equation}
where we introduced the field \( \varphi = v /(\pi \alpha') \) and used the definition of the gauge coupling for the \( p \)--brane effective action \( g_p^2 = \left(2\pi \right)^{p-2} \left(\alpha' \right)^{(p-3)/2} \). The indices are raised and lowered with the (undeformed) flat metric and repeated indices are summed over.
In a more compact notation, the action can also be written as a sum of squares,
\begin{equation}
  \mathscr{L}_{\epsilon_1, \epsilon_2} = \frac{1}{4 g_4^2} \Big( 1 + \| F \|^2 + \frac{1}{2} \| \di \varphi +  \im \epsilon \, \imath_{\hat U} F \|^2  + \frac{\epsilon^2}{8} \| \imath_{\hat U} \di ( \varphi + \bar \varphi) \|^2 \Big) \, ,
\end{equation}
where \( \imath \) is the interior product and
\begin{equation}
  \| V \|^2 = V \wedge * \bar V \, .  
\end{equation}
This is the form of the action for the \( \Omega \)--background that was discussed in~\cite{Nekrasov:2003rj,Nekrasov:2009rc}. 
Since \( \hat U \) depends explicitly on all the coordinates on the \D4--brane, Poincaré invariance is completely broken and the system is effectively zero-dimensional, but preserves two supersymmetries. 

\bigskip

Some of the contributions to the brane action come from the B--field, some
from the bulk dilaton and metric.  Let us point out the
latter first, as they are larger at small deformation, and additionally they are
odd under a certain discrete symmetry.
\begin{description}
\item[$B$--field couplings.] %
The cross terms with a single gauge field strength and a single scalar gradient 
are odd under the charge conjugation symmetry $A\lll\mu \to -A\lll\mu$.  This symmetry
is the same symmetry under which the bulk Neveu--Schwarz $B$--field is odd, and indeed
these terms in the brane action are induced by the first-order couplings in the 
\ac{dbi} action to $B\lll{\mu\n}$.  These terms are leading order in the deformation
parameter(s) $\epsilon$.
\item[Dilaton and metric deformation.] %
These deformations are of order $\epsilon^2$ and smaller, and invariant under all global
symmetries (other than the Poincar\'e group).  These terms control the classical properties
of gauge field configurations, such as instanton and multi-instanton configurations
and thus contribute directly to the deformation of the integrand on instanton moduli
spaces.
\end{description}

\bigskip

It is instructive to see how the deformation lifts the zero modes of \emph{e.g.} the
one-instanton solution.  We can see this from two complementary points
of view: the string-theoretic description of the instanton as a \D{}--instanton in the
presence of a \D3--brane; and the field-theoretic description of the instanton as a
low-energy object described as a gauge-field profile.

At the string theoretic level, a single pointlike \D{}--instanton couples only to the dilaton, not to
the metric and $B$--field.  The coupling to the latter two comes only through the fundamental
\D3--\D{(-1)} strings, whose condensation comprises a nonzero size for the instanton.  But
the dilaton couples to the \D{(-1)}--brane at any size, as one can see from the full
\ac{dbi} action for the \D{}--instanton:
\bbb
S\lll{\D{(-1)}} = 2\pi \exp [-\Phi ] \, .
\eee
The only critical point for the translational zero mode of the pointlike
\D{}--instanton is thus a critical point of the dilaton profile $\Phi(\mathbf{x})$.   In 
the fluxtrap solution (\ref{eq:IIA-TwoEpsilonDilaton}), the only critical point of $\Phi$ is
the fixed point $\rho_i = 0$ of the $U(1)$ action rotating the complex coordinates 
of $\IC\uu 2$.  The localization of the integral over instanton moduli space at
fixed points of the $U(1)$ action is implemented simply by the effective potential
induced by the dilaton.

Without referring directly to the string-theoretic origin of the action \eqref{eq:Two-epsilon-gauge},
we still infer the effective potential for a small instanton from the spatial dependence of
the quadratic action for the gauge connection.  Intuitively, an
instanton should seek the maximum of the gauge coupling, since its action goes 
as $g\lll 4 \uu{-2}$.  For an action such as the one in Eq.~\eqref{eq:Two-epsilon-gauge},
that is not Lorentz-invariant, there is not a uniquely defined ``gauge coupling'', since
the tensor defining the gauge kinetic term is not diagonal.  However a
sufficiently small instanton -- smaller than the typical scale
of variation $\e\uu{-1}$ of the couplings -- is a spherically symmetric pointlike object, and cannot sense the symmetry breaking.  Therefore it can only couple to the trace of the gauge kinetic tensor,
which in our case is 
\bbb
\Tr \left( \text{gauge kinetic tensor} \right)  = \frac{1}{g_4^2} \left( 1 + \tfrac{1}{2} \epsilon^2 \| U \|^2 \right) \equiv \frac{1}{g^2_{4,\text{scalar}}} \, , 
\eee
which upon comparison with the string solution in Eq.~\eqref{eq:general-fluxtrap}
does in fact turn out to equal $g_4^{-2} \exp[- \Phi]$, up to terms of order $\mathcal{O}(\epsilon^4 \|U\|^4)$.

\subsection{The \texorpdfstring{$\epsilon_1=-\epsilon_2 $}{opposite-epsilons} limit}
\label{sec:topol-string-limit}

\paragraph{Closed strings.}

Let us now consider the case of two identifications in flat space by taking the limit \( \epsilon_3 = 0 \) in the expressions in Eq.~(\ref{eq:IIA-TwoEpsilonBulk}). Now \( U \) is the Killing vector corresponding to the rotations in the \( (x^0, x^1) \) and \( (x^2, x^3 ) \) planes,   
\begin{equation}
  \textstyle  \epsilon\, U = \epsilon_1 \left( x^0 \del_1 - x^1 \del_0 \right) + \epsilon_2 \left( x^2 \del_3 - x^3 \del_2 \right) \, ,
\end{equation}
and the background reads
\begin{subequations}
  \label{eq:IIA-OneEpsilonTopologicalBulk}
  \begin{align}
    \di s^2 &= \di \mathbf{x}_{0\dots8}^2 + \frac{ (\di x^9)^2 - \left( \epsilon_1 \rho_1^2 \di \phi_1 + \epsilon_2 \rho_2^2 \di \phi_2 \right)^2}{ 1 + \epsilon_1^2 \rho_1^2 + \epsilon_2^2 \rho_2^2  } \, , \\
    B &=  \frac{\left( \epsilon_1 \rho_1^2 \di \phi_1 + \epsilon_2 \rho_2^2 \di \phi_2 \right) \wedge \di x^9}{ 1 + \epsilon_1^2 \rho_1^2 + \epsilon_2^2 \rho_2^2 } \, ,\\
    \eu^{-\Phi} &=  \sqrt{ 1 + \epsilon_1^2 \rho_1^2 + \epsilon_2^2 \rho_2^2 } \, .
  \end{align}
\end{subequations}
In order to preserve supersymmetry we impose
\begin{equation}
  \epsilon_1 = - \epsilon_2 = \epsilon\,,
\end{equation}
and we obtain \( 32/2 = 16 \) supercharges corresponding to the following Killing spinors:
\begin{equation}
  \begin{cases}
    \etaA^L =  \left(\Id + \Gamma_{11} \right) \exp [ \frac{\phi_1}{2} \gamma_{01} ] \exp [ \frac{\phi_2}{2} \gamma_{23} ] \exp [ \frac{\phi_3}{2} \gamma_{45} ] \left(\gamma_{01} + \gamma_{23} \right) \eta_0 \, ,  \\
    \etaA^R =  \left(\Id - \Gamma_{11} \right) \Gamma_u \exp [ \frac{\phi_1}{2} \gamma_{01} ] \exp [ \frac{\phi_2}{2} \gamma_{23} ] \exp [ \frac{\phi_3}{2} \gamma_{45} ] \left(\gamma_{01} + \gamma_{23} \right) \eta_1 \, ,  
  \end{cases}
\end{equation}
where \( \eta_0 \) and \( \eta_1 \) are constant real spinors and
\begin{align}
  x^0 + \im x^1 &= \rho_1 \eu^{\im \phi_1} \, , &
  x^2 + \im x^3 &= \rho_2 \eu^{\im \phi_2} \, .
\end{align}
This is the fluxtrap background introduced in~\cite{Hellerman:2011mv}.

\paragraph{Open strings.}

If we introduce an \NS5--\D4 system as in the previous case (see Table~\ref{tab:D4-embedding-two-epsilon}), we obtain a configuration that preserves four supersymmetries.
The action is formally the same as in Eq.~(\ref{eq:Two-epsilon-gauge}), but this time the pullback of the Killing vector \( U \) is
\begin{equation}
  \hat U = f^* U =  \xi^0 \partial_1 - \xi^1 \partial_0 - \xi^2 \partial_3 + \xi^3 \partial_2 \,  .
\end{equation}
Poincaré invariance is completely broken also in this case, but the system has four supercharges.

The instanton partition function for this four-dimensional theory is identified 
with the field theory limit of the topological string partition function with coupling \( \gtop \propto \epsilon \)~\cite{Nekrasov:2002qd,Iqbal:2003ix}.
 In this context \( \epsilon \) is the coupling of the graviphoton field in the gauge theory obtained by reducing M--theory on the Melvin circle~\cite{Antoniadis:1993ze}. The resulting Ramond--Ramond \tIIA background provides in this sense a different realization of the $\Omega$--deformation~\cite{Billo:2006jm}. A more detailed discussion of the relationship between our construction and topological strings is presented in Section~\ref{sec:topstring}.

\subsection{The \acl{ns} limit $ \epsilon_1 = 0$}\label{sec:NS}

Another remarkable limit of the bulk fields in Eq.~(\ref{eq:IIA-TwoEpsilonBulk}), is given by \( \epsilon_1  = 0\). This time we impose \( \epsilon = \epsilon_2 = -\epsilon_3 \) and the resulting background has \( 32/2 = 16 \) supercharges.

Once more we look at the effective theory for a \D4--brane suspended between two \NS5--branes as in Table~\ref{tab:D4-embedding-two-epsilon}. The configuration preserves four supercharges.
The \textsc{dbi} action is still formally the same:
\begin{equation}
  \mathscr{L}_{\epsilon} = \frac{1}{4 g_4^2} \left[ 1 + F_{ij} F^{ij} + \frac{1}{2} \left( \partial_i \varphi + \im \epsilon \hat U^k F_{ki}  \right) \delta^{ij}  \left( \partial_j \bar \varphi - \im \epsilon \hat U^l F_{lj} \right) 
  - \frac{\epsilon^2}{8} \left( \hat U^i \partial_i ( \varphi + \bar \varphi)  \right)^2  \right] \, ,
\end{equation}
where the pullback of the vector \( U \) is
\begin{equation}
  \hat U = f^* U =  - \xi^3 \del_2 + \xi^2 \del_3  \,  .
\end{equation}
In this case, the Poincaré invariance is only broken in the directions \( \xi^2 \) and \( \xi^3 \) and the system is effectively an \( \mathcal{N} = (2,2) \) two-dimensional gauge theory. This is precisely the action discussed by Nekrasov and Shatashvili in~\cite{Nekrasov:2009rc}.

This type of four dimensional gauge theory is related to the quantization of integrable models~\cite{Nekrasov:2009rc,Aganagic:2011mi}. Starting from a four-dimensional $\mathcal{N}=2$ \ac{sw} gauge theory subjected to the $\Omega$--background with $\epsilon_1=0$ discussed in this section, one obtains $\mathcal{N}=2$ super-Poincaré invariance in two dimensions. The crucial observation of~\cite{Nekrasov:2009rc} is that 
the two-dimensional twisted superpotential derived from the prepotential in the four-dimensional theory can be identified with the Yang--Yang counting function of a quantum integrable system. The supersymmetric vacua are mapped to the eigenstates of a \emph{quantum} integrable system whose Planck constant is given by the deformation parameter $\epsilon_2 = \hbar$.  For $\epsilon_2\to0$ one recovers the classical integrable system whose spectral curve is given by the \ac{sw} curve.

We will show in the following (Section~\ref{sec:non-commutative-ns-limit}) that Neveu--Schwarz $B$--field resulting from the Melvin deformation of our background geometry give rise precisely to the type of non-commutativity of the spectral curve that one expects from the corresponding quantum integrable model.

\section{M--theory lift}
\label{sec:m-theory-lift}
The theories that we are discussing can be understood in terms of deformations of four-dimensional \( \mathcal{N} = 2  \) \ac{sw} theories. It is natural to describe them in an M--theory setting, following~\cite{Witten:1997sc}.
In this section we will show how the fluxtrap construction lifts to eleven dimensions and how the \( \Omega \)--deformation affects the dynamics of the \M5--branes that realize the gauge theory.

\subsection{The Bulk}
\label{sec:bulk}

A \tIIA background with metric \( g \), Neveu--Schwarz field \( B \), dilaton \( \Phi \), one-form \( C_1 \) and three-form \( C_3 \) is oxidized on a circle \( x^{10} \) to M--theory with metric \( G \) and three-form \( A_3 \) as follows~\cite{Polchinski:1998rr}:
\begin{align}
  G_{IJ} \di x^I \di x^J &= \eu^{-2\Phi/3} g_{ij} \di x^i \di x^j + \eu^{4\Phi/3} \left(\di x^{10} + A_1 \right)^2 \, ,\\
  C_3 &= A_3 + B \wedge \di x^{10} \, .
\end{align}
Starting from the \tIIA background in Eq.~(\ref{eq:general-fluxtrap}) we find the general form of the \emph{M--theory fluxtrap background}:
\begin{subequations}
  \label{eq:M-fluxbrane}
  \begin{align}
    G_{IJ}\di x^I \di x^J &= \left( 1 + \epsilon^2 \|U\|^2 \right)^{1/3} \left[ g_{ij}\di x^i\di x^j + \frac{ (\di x^9)^2 + (\di x^{10})^2 - \epsilon^2 U_i U_j \di x^i \di x^j }{ 1 + \epsilon^2 \|U\|^2 }\right] \, ,\\
    C_3 &= \epsilon\,  \frac{U_i\di x^i\wedge \di x^9 \wedge\di x^{10}}{1 + \epsilon^2 \|U\|^2} \, .
  \end{align}
\end{subequations}
It is interesting to remark that the directions \( x^9 \) and \( x^{10} \) which have completely different origins (\( x^9 \) is the dual of the Melvin circle while \( x^{10} \) is the M--circle) enter the background in a completely symmetric fashion.

In the following we will study the embedding of an \M5--brane in this background. For this purpose it is interesting to consider the physics close to the center of the fluxtrap, \emph{i.e.} the limit \( \epsilon^2 \|U\|^2 \ll 1 \). The fields become
\begin{subequations}
  \begin{align}
    G_{IJ} \di x^I \di x^J &= g_{ij}\di x^i\di x^j + (\di x^9)^2 + (\di x^{10})^2 + \mathcal{O}(\epsilon^2 \| U \|^2)\, ,\\
    C_3 &= \epsilon \, U_i\di x^i\wedge \di x^9 \wedge\di x^{10} + \mathcal{O}(\epsilon^3 \| U \|^3)\, .
  \end{align}
\end{subequations}
The appropriate setting to discuss the gauge theories found in the previous section is obtained by starting from a flat metric \( g_{ij} = \delta_{ij} \). In this case, at this order the metric is flat and there is a constant four-form flux \( F_4 = \di A_3 \),
\begin{subequations}
  \label{eq:first-order-Mbulk}
  \begin{align}
    G_{IJ} &= \delta_{IJ} + \mathcal{O}(\epsilon^2 \| U \|^2) \, ,\\
    F_4 &= 2\, \epsilon \, \omega \wedge \di x^9 \wedge\di x^{10} + \mathcal{O}(\epsilon^3 \| U \|^3) \, ,
  \end{align}
\end{subequations}
where \( \omega \) is the linear combination of the volume forms of the planes in which the original Melvin identifications have been performed,
\begin{equation}
  \di [\epsilon \, U_i \di x^i ] = 2 \epsilon \, \omega \equiv 2 \sum_{k=1}^N \epsilon_k \omega_k \, .  
\end{equation}
Note that \( \omega \) is the graviphoton field strength in~\cite{Nekrasov:2002qd}.

\subsection{M5--brane embedding}
\label{sec:m5-embedding}

Having constructed the M--theory background we are now ready to study the embedding of \M5--branes. It is known that a configuration of \D4--branes suspended between \NS5--branes in flat space as in Table~\ref{tab:D4-embedding-two-epsilon} lifts to a single \M5--brane wrapped on a Riemann surface~\cite{Witten:1997sc}. We want to see how the presence of the fluxbrane modifies this picture.

\begin{table}
  \centering
  \begin{tabular}{lccccccccccc}
    \toprule
    \( x \)   & 0               & 1               & 2               & 3  & 4            & 5  & 6      & 7 & 8  & 9 & 10 \\
    \midrule
    fluxbrane & \ep{\epsilon_1} & \ep{\epsilon_2} & \ep{\epsilon_3} & \X & \X           & \X & \T     & \T              \\
    \NS5      & \X              & \X              & \X              & \X &              &    &        &   & \X & \X     \\
    \D4       & \X              & \X              & \X              & \X &              &    & \X     &   &    &   & \X \\
    \midrule
    \( \xi \) & \ep{u}          & \ep{w}          &                 &    & \( \Re(s) \) &    & \ep{v} & \(\Im(s)\)      \\
    \bottomrule
  \end{tabular}
  \caption{The embedding of the \M5--branes resulting from the lift of the \NS5 and \D4 in Section~\ref{sec:omega-def}. The directions \( x^9 \) and \( x^{10} \) enter symmetrically in the background. The bottom line contains the complex coordinates used for the description.}
  \label{tab:M5-embedding-two-epsilon}
\end{table}

The simplest approach consists in looking for the most general \M5--brane preserving the same supersymmetries as the \NS5--\D4 system~\cite{Howe:1997ue}. The \ac{bps} condition can be expressed in terms of a projector~\cite{Howe:1997fb},
\begin{equation}
  \proj{\M5}_+ = \tfrac{1}{2}\left( \Id + \Gamma^{\M5} \right) \etaM = 0\,,
\end{equation}
where \( \etaM \) is the generic Killing spinor preserved by the background and \( \Gamma^{\M5} \) is
\begin{align}
  \Gamma^{\M5} &= \left( - \Id + \frac{1}{3} \hat \Gamma^{m_1 m_2 m_3} h_{m_1 m_2 m_3} \right) \Gamma_{(0)} \, ,&
  \Gamma_{(0)} &= \frac{1}{6! \sqrt{- \hat g}} \eta^{m_1 \dots m_6} \hat \Gamma_{m_1 \dots m_6}\,,
\end{align}
and \( \hat \Gamma \) and \( \hat g \)  are respectively the pullbacks of the gamma matrices and the metric to the brane.\footnote{The embedding is defined by a map
\begin{equation*}
  \begin{aligned}
    f : \M5 &\to \text{bulk}\,, \\
     \zeta^m & \mapsto x^I (\zeta^m)\,.
  \end{aligned}
\end{equation*}
For a given vielbein \( \EM\indices{^A_I} \) for the bulk metric we define
\begin{equation*}
  \hat e^A = f^* \EM^A = \hat e\indices{^A_m} \di \zeta^m = \EM\indices{^A_I} \del_m x^I \di \zeta^m\,,
\end{equation*}
and the pullbacks \( \hat g  \) and \( \hat \Gamma \) are given by
\begin{align*}
  \hat g &= f^* g = \hat g_{mn} \di \zeta^m \di \zeta^n = G_{IJ} \del_m x^I \del_n x^J \di \zeta^m \di \zeta^n = \hat e^a \hat e^B \delta_{AB}\, ,\\
  \hat \Gamma_m &= \gamma_A \hat e\indices{^A_m} = \gamma_A \EM\indices{^A_I} \del_m x^I \, ,
\end{align*}
where \( \gamma_A \) are the flat gamma matrices in eleven dimensions.} The other degrees of freedom of the \M5--brane are represented by a selfdual three-form \( h \),
\begin{equation}
  h = * h \,, 
\end{equation}
which is related in a non-linear way to the pullback of the bulk four-form flux \( F_4 \), \emph{viz}.
\begin{gather}
  \di \hat H^{[3]}  = - \frac{1}{4} f^* F_4 \, , \\
  \begin{aligned}
    \hat H^{[3]}_{mnp} &= m\indices{_m^q} m\indices{_n^r} h_{p q r} \,
    , &
    m\indices{_m^n} &= \delta\indices{_m^n} - 2 h_{mpq} h^{npq} \, .
  \end{aligned} 
\end{gather}

Since we want to describe the lift of the generic \( \Omega \)--deformed four-dimensional theory with \( \epsilon_1 \neq \epsilon_2 \) we start from the \tIIA background in Eq.~(\ref{eq:IIA-TwoEpsilonBulk}) and lift it to eleven dimensions as in Eq.~\eqref{eq:M-fluxbrane}. In our conventions the eleven-dimensional Killing spinors \( \etaM \)  are related to the ten-dimensional ones \( \etaA \) by 
\begin{equation}
  \etaM = \eu^{-\Phi/6} \etaA \, ,  
\end{equation}
where \( \Phi \) is the \tIIA dilaton in Eq.~(\ref{eq:IIA-TwoEpsilonDilaton}) and \( \etaA \) are the Killing spinors in Eq.~(\ref{eq:IIA-TwoEpsilonSpinors}) (see Appendix~\ref{sec:supersymm-conv}).

\bigskip 

As a first step let us find the supersymmetries preserved separately by the lifts of the \NS5 and \D4--branes (see Table~\ref{tab:M5-embedding-two-epsilon}).
\begin{description}
\item[The \emph{\NS5}--brane] is lifted to an \M5--brane extended in \( (x^0, \dots, x^3, x^8, x^9) \). The pullback of the four-form flux vanishes \( f^*_{\NS5} F_4 = 0\) and the kappa symmetry projector is
  \begin{equation}
    \proj{\NS5}_+ = \tfrac{1}{2} \left( \Id + \gamma_{012389} \right) \, ;  
  \end{equation}
\item[The \emph{\D4}--brane] is lifted to an \M5--brane extended in \( (x^0, \dots, x^3, x^6, x^{10}) \). Also in this case the pullback of the four-form flux vanishes, but we need to take into account the deformed metric. The result is that the kappa symmetry projector reads
  \begin{equation}
    \proj{\D4}_+ = \frac{1}{2} \left ( \Id + \frac{\gamma_{026} \left(\gamma_{13} + \epsilon_1 \abs{u} \gamma_{39} + \epsilon_2 \abs{w} \gamma_{19} \right) \gamma_{10}}{\sqrt{1+ \epsilon_1^2 \abs{u}^2 + \epsilon_2^2 \abs{w}^2}} \right),
  \end{equation}
  where \( u = x^0 + \im x^1 \) and \( w = x^2 + \im x^3 \). 
\end{description}
The kappa~symmetry projectors for the \NS5 and \D4 commute,
\begin{equation}
  \left[ \proj{\NS5}_+, \proj{\D4}_+ \right] = 0 \, ,
\end{equation}
and each breaks half of the supersymmetries. As already observed in Section~\ref{sec:generaleps} we are looking for embeddings preserving one sixteenth of the thirty-two Killing spinors of e\-lev\-en-di\-men\-sion\-al supergravity (one quarter from the $\epsilon$--deformation in the bulk and one half for each brane).

It is convenient to introduce complex coordinates in the bulk,
\begin{equation}
  \begin{cases}
    v = x^8 + \im x^9\,, \\
    s = x^6 + \im x^{10}\,,
  \end{cases}
\end{equation}
and make the following ansatz for the embedding of the lifted \NS5--\D4 system:
\begin{equation}
  f_{\M5} : (u,z,w) \mapsto
  \begin{cases}
    x^0 + \im x^1 = u\,, \\ x^2 + \im x^3 = w\,, \\ s = s(z, \bar z)\,, \\ v = v (z,\bar z)\,,
  \end{cases}
\end{equation}
where \( (u,w,z) \) are complex local coordinates on the brane.

Since we are interested in the physics in a neighborhood of the fluxtrap we can expand in powers of \( \epsilon \). The embedding expanded at linear order is
\begin{equation}
  \begin{cases}
    s (z, \bar z) = s_0 (z, \bar z) + \epsilon \, s_1 (z, \bar  z) + \dots \\
    v (z, \bar z) = v_0 (z, \bar z) + \epsilon \, v_1 (z, \bar  z) + \dots     
  \end{cases}
\end{equation}
For \( \epsilon = 0 \) we are back to the standard configuration of flat space without four-form flux. In this case the \ac{bps} condition to solve is
\begin{equation}
  \proj{\M5}_+ \proj{\NS5}_- \proj{\D4}_- \etaM = 0 \, .  
\end{equation}
This is satisfied if both \( s_0 \) and \( v_0 \) are holomorphic functions of \( z \),
\begin{equation}
  \begin{cases}
    s_0 = s_0 (z)\,, \\ v_0 = v_0 (z) \, .
  \end{cases}
\end{equation}
In other words, the \M5--brane is wrapped on a Riemann surface $\Sigma$ in the \( \setC^2 \) plane generated by \( s \) and \( v \).

At first order in \( \epsilon \) we want to discuss the embedding
\begin{equation}
  \begin{cases}
    s (z, \bar z) = s_0 (z) + \epsilon \, s_1 (z, \bar  z) + \dots \\
    v (z, \bar z) = v_0 (z) + \epsilon \, v_1 (z, \bar  z) + \dots     
  \end{cases}
\end{equation}
in the background of Eq.~(\ref{eq:first-order-Mbulk}). The four-form flux has a non-vanishing pullback on the brane coming from the \( (1,1) \) component in the \( s,v  \) plane:
\begin{equation}
  f_{\M5}^* F_4 = \frac{\im}{4} \left(  \bdel \bar s_0 \del v_0 - \del s \bdel  \bar v_0 \right) \di z \wedge \di \bar z \wedge \left( \epsilon_1 \di u \wedge \di \bar u + \epsilon_2 \di w \wedge \di \bar w \right) \, .
\end{equation}
Note that the pullback only depends on the embedding at the next-lowest order in $\epsilon$. The expression can be suggestively  recast in the form
\begin{equation}
\label{eq:M5-F4-pullback}
  f_{\M5}^* F_4 = \epsilon \, \omega_\Sigma \wedge ( f^*_{\M5} \omega ) \, ,  
\end{equation}
where \( \omega \) is again the weighted sum over the planes of the Melvin identifications and \( \omega_\Sigma \) is the volume form of the Riemann surface $\Sigma$ with Kähler potential
\begin{equation}
  K(z,\bar z) = \frac{1}{8} \Im ( v_0(z) \bar s_0(\bar z) )\,.  
\end{equation}
The corresponding metric and volume form are:
\begin{align}
  \di s^2 &= 4 \del \bdel K (z,\bar z) \di z \di \bar z\,, & \omega_\Sigma &= \frac{\bdel \bar s_0 \del v_0 - \del s_0 \bdel \bar v_0 }{2} \di z \wedge \di \bar z = 8 \im \del \bdel K \di z \wedge \di \bar z\, .
\end{align}

Since \( f^*_{\M5} F_4 \) has only first-order terms in \( \epsilon \), the selfdual three-form \( h \) on the brane obeys a simple linear condition,
\begin{equation}
  \di h = - \frac{1}{4} f^*_{\M5} F_4 \, .
\end{equation}
Self-duality naturally breaks \( h \) into two pieces\footnote{Algebraically this corresponds to the identification \( \mathfrak{so(4)} \simeq \mathfrak{su}(2) \oplus \mathfrak{su}(2) \) under which \( \epsilon_1 \) and \( \epsilon_2 \) are mapped to \( \epsilon_+ \) and \( \epsilon_- \).}:
\begin{multline}
  h = \frac{1}{2} \big( \left( \epsilon_2 - \epsilon_1 \right) \del K(z,\bar z) \di z \wedge \left( \di u \wedge \di \bar u - \di w \wedge \di \bar w \right) \\ + \left( \epsilon_2 + \epsilon_1 \right) \bdel K(z,\bar z) \di \bar z \wedge \left( \di u \wedge \di \bar u + \di w \wedge \di \bar w \right) \big),
\end{multline}
and is more conveniently written in terms of \( \epsilon_{\pm} = \epsilon_2 \pm \epsilon_1 \):
\begin{equation}
  \boxed{%
    h = - \im \left(\epsilon_- \del K \di z \wedge (f^*_{\M5} \omega_-) + \epsilon_+ \bdel K \di \bar z \wedge (f^*_{\M5} \omega_+) \right) }  \, .
\end{equation}
Having found $h$, one can now evaluate explicitly
\begin{equation}
  \Gamma^{\M5}_{\mathcal O(\epsilon)} = \left( - \Id + \tfrac{1}{3} h_{m_1m_2m_3} \hat\Gamma^{m_1m_2m_3} \right) \Gamma_{(0)} \, ,
\end{equation}
and impose the kappa~symmetry projection
\begin{equation}
    \proj{\M5}_+ \proj{\NS5}_- \proj{\D4}_- \etaM = 0 \,, 
\end{equation}
which is greatly simplified by the fact that the contribution of the $h$ field is projected out by supersymmetry
\begin{equation}
  h_{m_1m_2m_3} \hat\Gamma^{m_1m_2m_3}\Gamma_{(0)} \proj{\NS5}_- \proj{\D4}_- \equiv 0 \, .
\end{equation}
This means that we are back to
\begin{equation}
  \left( \Id - \Gamma_{(0)} \right) \proj{\NS5}_- \proj{\D4}_- \etaM = 0\,,
\end{equation}
which is precisely the same equation as in the $\epsilon=0$ case and is satisfied by the same Cauchy--Riemann conditions
\begin{equation}
  \begin{cases}
    s_1 = s_1 (z) \,,\\
    v_1 = v_1 (z)\,.
  \end{cases}
\end{equation}
At this order, the embedding is thus still a Riemann surface, but this time a non-vanishing self-dual three-form flux is turned on on the M5--brane.

\bigskip

At first order in \( \epsilon \), the \( (2,0) \) theory on the worldvolume of the \M5--brane can still be decomposed into a product of a four dimensional part \( M_4\) and a Riemann surface \( \Sigma \). 
Turning on the \( \Omega \)--deformation on \( M_4 \) automatically gives rise to a flux on the whole \M5--brane. In particular, the flux has also non-vanishing components on \( \Sigma \),  where it acts as a Kähler form.

It is an important task for the future to extend the above treatment to quadratic order in $\epsilon$ at which instanton localization takes place as we have seen in Section~\ref{sec:generaleps}. It is likely that beyond linear order, the surface $\Sigma$ may no longer be holomorphic.

\subsection{Relationship with the topological string}\label{sec:topstring}

Much has been said about the relationship of the topological string
to the $\Omega$--de\-for\-ma\-tion of gauge theory (\emph{e.g.}~\cite{Iqbal:2003ix, Dijkgraaf:2007sw,Nekrasov:2010ka}).  By realizing the four-dimensional gauge theory as the dynamics of
a geometrically engineered Calabi--Yau singularity in \tIIA string theory, the gauge theory 
can be lifted to a five-dimensional theory living on the same singularity in M--theory,
and then re-compactified with Melvin boundary conditions to yield an $\Omega$--deformed
gauge theory in four dimensions, which is ultraviolet-completed to the topological
string on the Calabi--Yau singularity.  Here, the topological string coupling~$\gtop$ is
directly proportional to the parameter $\e = \e\lll 1 = - \e\lll 2$ of the Melvin twist.  This logic
has been verified quantitatively in a number of examples ~\cite{Iqbal:2003ix}.

It is interesting to compare our solution to the topological string realization
of the same deformation of the same $\mathcal{N} = 2$ gauge theory.
We have realized the $\Omega$--deformed $\mathcal{N} = 2$ gauge dynamics as 
the dynamics of a $(p,q)$ fivebrane web~\cite{Aharony:1997bh} compactified
on a circle with Melvin boundary conditions; the topological string
realizes the deformed gauge theory \emph{via} the same construction, except with the
replacement of the $(p,q)$ fivebrane web with a local Calabi--Yau geometry in eleven-dimensional M--theory as the origin of the five-dimensional gauge theory.  

In our realization, there exists
a limit in which all degrees of freedom of the gauge theory are realized as open strings 
on \D{}--branes living in a spacetime with only Neveu--Schwarz background fields turned on; 
this
makes it possible write an explicit action deformed gauge theory.  In principle,
the existence of a well-defined perturbative string realization makes it
possible to include string and five-dimensional Kaluza--Klein corrections
to the renormalizable four-dimensional action, if desired.

Prior to reduction on the Melvin circle, the two five-dimensional theories are
not the same and have different properties.  In particular, the R--symmetry groups are different in
the two five-dimensional theories, beyond the low-energy level.  In the $(p,q)$ fivebrane
web, as noted earlier, there is an exact $SU(2)$ R--symmetry rotating three common
transverse coordinates to all the branes.  In the non-compact Calabi--Yau singularities
of interest for the study of the topological string, the R--symmetry is generically only a 
$U(1)$.

Both our construction and that of the topological string can be followed through a set of mutually equivalent dual frames, among which in each case is an M--theory solution with \M5--branes and 
four-form flux.  In this last description, the R--symmetry of the solution representing the topological string is enhanced from $U(1)$ to $SU(2)$ as a third noncompact transverse direction decompactifies.
This description realizes the gauge theory degrees of freedom as coming from the six-dimensional $(0,2)$
superconformal theory on a Riemann surface as in \cite{Witten:1997sc}, deformed by the
presence of four-form M--theory flux with various numbers of components longitudinal
and transverse to the \M5--branes.  In both cases, the four-form flux
is proportional to the parameter $\e$ deforming the gauge theory.  In the limit
$\e\to 0$, both the \M5--brane dual frame of our construction and the \M5--brane dual frame of the topological string theory are of exactly the same type: a set of fivebranes wrapping a Riemann surface in a flat eleven-dimensional background.

At leading order in $\e$ level the deformations are subtly different.
In particular, the type of four-form flux differs between
the two solutions.  In both configurations, the flux is written as the product of two-forms (see Eq.~\eqref{eq:first-order-Mbulk} and \emph{e.g.} Equation~(3.19) of~\cite{Dijkgraaf:2007sw}):
\newcommand*{\wA}{\ensuremath{\omega_{\setR^4}}}
\newcommand*{\wB}{\ensuremath{\omega_{\setC^2}}}
\bbb
F_{4} = 
\wA \wedge \wB \, .
\eee
In both cases, the first factor is the same self-dual
harmonic form describing the effective graviphoton flux in the directions $x^{0,1,2,3} = \setR^4$
(or \ac{tn}).
The second
factor $x^{6,8,9,10} = \setC^2$ has a complex structure with respect to which the Riemann surface is
embedded holomorphically.  With respect to that complex structure, the flux
\( \wB \) in~\cite{Dijkgraaf:2007sw} has Hodge numbers~$(2,0)$ and~$(0,2)$ only, and consequently has vanishing integral on the Riemann surface.  In the solutions
presented here, the flux \( \wB \)  has a component with
Hodge numbers~$(1,1)$ and generically has a nonzero integral along the Riemann
surface (see Eq.~\eqref{eq:M5-F4-pullback}).

The two types of flux deformation induce the same term in the deformation of the gauge theory action at order $\mathcal{O}(\e)$, except with a different linear combination of the
scalars appearing.   In the \NS5/\D4 construction
with weak $g\lll s$, the \D4--branes are parallel segments pointing in the $x\uu 6$ direction between \NS5--branes, with a gauge 
group and adjoint degrees of freedom living on the stack of parallel branes in the
segment between each pair of adjacent \NS5's.  Our $\Omega$--deformation of
the background \eqref{eq:IIA-TwoEpsilonBulk} affects each set of gauge and adjoint degrees of freedom in the same way, proportionally to the inverse gauge coupling of each gauge group:
from \eqref{eq:Two-epsilon-gauge} we see that at small $\e$ the deformation contributes 
\bbb
\Delta \mathscr {L} = \frac{\e}{4 g\lll 4\sqd}  \hat{U} \uu k \Tr \big (  F\lll{ki} \nabla\uu i {\Im}(\varphi) \big )
+ \mathcal{O} (\e\sqd )\ .
\label{OurFirstOrderGaugeDeformation}
\eee

The $\Omega$--deformation implemented by four-form flux in the \M5--brane duality frame
of the topological string induces a term of a similar form, through a different
orientation of the flux on the same type of Riemann surface.  The four-form
flux deformation there (see \emph{e.g.}~\cite{Dijkgraaf:2007sw}) is proportional to
\bbb
\Delta F\lll 4 \propto \wTN \wedge \di s \wedge \di v + \text{ c.c.} \, ,
\eee
where \( \wTN \) is the \( U(1) \)--invariant two-form on the \ac{tn} space (see Equation~\eqref{eq:omega-TN}).  
Reducing on $x^{10} = \Im (s)$ to \tIIA, the M--theory four-form flux 
contributes to the NS/NS three-form flux as  $\Delta H \propto  \wTN \wedge \di x^9$, and
to the Ramond--Ramond four-form flux as
\bbb
\Delta F\uu{\textsc{iia}}\lll 4 \propto \wTN \wedge \di x^6 \wedge \di x^8 \ .
\eee
The NS flux contributes to the brane action exactly a term proportional to \eqref{OurFirstOrderGaugeDeformation}.
Through the Chern--Simons term on the \D4--branes, the Ramond--Ramond
flux induces a coupling
\bbb
\Delta S\lll{\D4} = \im \int \cc \wTN \wedge \di x \uu 6 \wedge 
\Tr ( \di x \uu 8 \wedge A )\ ,
\eee
where $A$ is the gauge connection on the \D4--brane.  As a four-dimensional action,
this generates a term proportional to $L\lll 5 = g\lll 4\sqd / g\lll 5\sqd$:
\bbb
\Delta S\lll{\text{ 4d}}  \ni \frac{\im}{g\lll 4\sqd} \int \wTN \wedge 
{\Tr} (\di x\uu 8 \wedge A )\ .
\eee
Integrating by parts and exploiting the fact in the case $\e\lll 2 = - \e\lll 1$ that 
\bbb
\wTN = * \wTN 
\propto \di (U\lll i \di x \uu i)\ ,
\eee
we have
\bbb
\Delta \mathscr{L} \ni \frac{1}{g\lll 4\sqd} \hat{U}\uu i \Tr (\nabla\uu k (x^8)   F\lll{ik} )\ ,
\eee
which we see is also proportional to \eqref{OurFirstOrderGaugeDeformation}, except 
involving the scalar $  {\Re}(\varphi)  \propto {\Re}(v) = x\uu 8 $ instead of
$   {\Im}(\varphi)  \propto  {\Im}(v)   = x\uu 9$.
It is not yet known whether our string embedding of the Nekrasov partition function
agrees with the topological string beyond the renormalizable level.
If so, it is tempting to think that our embedding may be related directly by some duality to the topological string.  Whatever the duality, it cannot be a duality that is realized as geometric in eleven dimensions, since the flux along the Riemann surface is a geometric invariant.   It is possible that the duality of~\cite{Leung:1997tw} between five-dimensional theories on Calabi--Yau singularities and five-dimensional theories on $(p,q)$~fivebrane webs may point towards the correct relationship, after compactification on a circle with Melvin--Nekrasov boundary conditions.

\section{9/11 flip and non-commutativity}\label{sec:9-11}

An equivalent realization for the gauge theory in the \( \Omega \)--background is obtained when compactifying the M--theory description on the circular orbits of the isometry \( \del_\phi \). This will lead naturally to a non-commutative structure which has already been associated to topological strings and the \( \Omega \)--deformation in the \ac{ns} limit~\cite{Nekrasov:2009rc,Aganagic:2003qj,Dijkgraaf:2007sw}.
While work has been done to develop a \ac{sw} map directly in M--theory (see e.g.~\cite{Chen:2010br}), we choose here to go the route of performing the usual \ac{sw} map in \tIIA string theory after a 9--11 flip.

In this section, we will not start out from a flat geometry, but from a \ac{tn} space, resulting in a \acs{ttrap} geometry. As already noted this is equivalent from the point of view of the quantities counted by Nekrasov's partition function. On the other hand, it is convenient in this situation because a $Q$--centered \ac{tn} space corresponds to $Q$ coincident \D6--branes in flat space in the right duality frame and our argument is most straightforward in the presence of a \D{}--brane with $B$--field. Moreover, the supersymmetry generators remain unchanged since this geometry preserves the very same supersymmetries as the fluxtrap in flat space (which is recovered as the $r\to 0$ limit of the \acs{ttrap}). Finally, in the limit $r\to\infty$ this background is a string theory realization of the alternative description of the $\Omega$--background proposed by Nekrasov and Witten~\cite{Nekrasov:2010ka}.

\subsection{The \texorpdfstring{$\epsilon\lll 1 = - \epsilon\lll 2$}{opposite-epsilons} limit}

As a first example let us consider a fluxtrap on a space of the form \( \mathrm{TN}_Q \times S^1 \times \setR^5 \). It is convenient to choose a coordinate system such that the initial metric (prior to identifications) is written as
\begin{equation}
  g_{ij} \di x^i \di x^j = V(r) \di \mathbf{r}^2 + \frac{1}{V(r)} \left( \di \theta + Q \cos \omega \di \psi \right)^2 + \di \mathbf{x}_{4 \dots 8}^2+ (\di \tilde x^9)^2  \, ,
\end{equation}
where
\begin{equation}
  \di \mathbf{r}^2 = \di r^2 + r^2 \di \omega^2 + r^2 \sin^2 \omega \di \psi^2 \, ,
\end{equation}
and
\begin{equation}
  V(r) = \frac{1}{\rTN^2} + \frac{Q}{r} \, .
\end{equation}
As shown in Appendix~\ref{sec:taub-nut-coordinates}, applying the following shifts to the Killing vector \( \partial_\theta \) corresponds to the case \( \epsilon_1 = - \epsilon_2 = \epsilon \):
\begin{align}
  \begin{cases}
    \tilde u \simeq  \tilde u + 2 \pi n_u\,, \\
    \theta \simeq \theta + 4 \pi \widetilde R \,\epsilon n_u\,.
  \end{cases} && n_u \in \setZ
\end{align}
 Introducing the disentangled variable \( \phi = \theta -2 \wt R \epsilon \tilde u \), we see that in this case,
\begin{align}
  U^i \del_i &= \del_\phi\,, & U_i \di x^i &= \frac{\di \phi + Q \cos \omega \di \psi}{V(r)}\,, & \|U\|^2 &= \frac{1}{V(r)}\,.
\end{align}
Note that in this case \( \del_\theta \) is a bounded isometry,
\begin{equation}
  \|U\|^2 = \frac{1}{V(r)} < \rTN^2  \,,
\end{equation}
which means that the fluxbrane does not break any additional supersymmetries\footnote{The fluxtrap construction on the isometry \( \del_\phi \) is natural in these coordinates because it uses the same \( U(1) \) fibration of the \ac{tn} seen as a hyperkähler quotient.}. 

Using the formula in Eq.~(\ref{eq:general-fluxtrap}) we find that the \acs{ttrap} background is given by
\begin{subequations}
  \begin{align}
    \di s^2 &= V(r) \di \mathbf{r}^2+ \frac{1}{V(r) + \epsilon^2} \left( \di \phi + Q \cos \omega \di \psi \right)^2 + \frac{V(r)}{V(r) + \epsilon^2} (\di x^9)^2 + \di \mathbf{x}^2_{4 \dots 8}\,, \\
    B &= \frac{\epsilon}{V(r) + \epsilon^2} \left( \di \phi + Q \cos \omega \di \psi \right) \wedge \di x^9 \,,\\
    \eu^{-\Phi} &=  \sqrt{ 1 + \frac{\epsilon^2}{V(r)}}\,.
  \end{align}
\end{subequations}
The corresponding Killing spinors can be found by using the general prescription of Section~\ref{sec:fluxtrap} and the expression for the Killing spinors of the undeformed \ac{tn} in Eq.~(\ref{eq:TN-spinors-GHcoord}). As expected, we find that the fluxtrap does \emph{not} break supersymmetry since the corresponding projector is the same that appears in the undeformed \ac{tn}. More explicitly, the \acs{ttrap} background preserves a total of \emph{sixteen supersymmetries}:
\begin{equation}
  \begin{cases}
    \etaA^L = \left( \Id + \Gamma_{11} \right) \exp[\frac{\omega}{2} \gamma_{01}] \exp[ \frac{\psi}{2} \gamma_{23} ] \left( \gamma_{01} + \gamma_{23} \right) \eta_0 \, , \\
    \etaA^R = \left( \Id - \Gamma_{11} \right) \Gamma_u \exp[\frac{\omega}{2} \gamma_{01}] \exp[ \frac{\psi}{2} \gamma_{23} ] \left( \gamma_{01} + \gamma_{23} \right) \eta_1 \, ,
  \end{cases}
\end{equation}
where
\begin{equation}
  \Gamma_u = \frac{- \epsilon \gamma_3 + \sqrt{V(r)} \gamma_9}{\sqrt{\epsilon^2 + V(r)}}\,.
\end{equation}
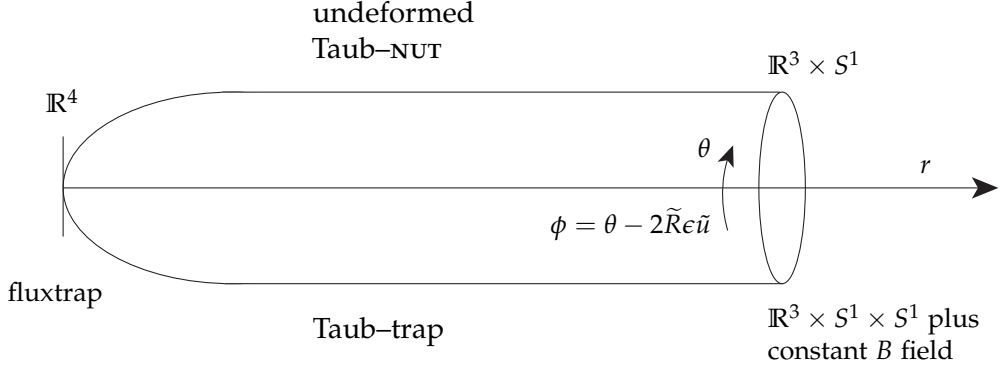
\begin{figure}
  \centering
    \begin{tikzpicture}
      \node (0,0) {\includegraphics[]{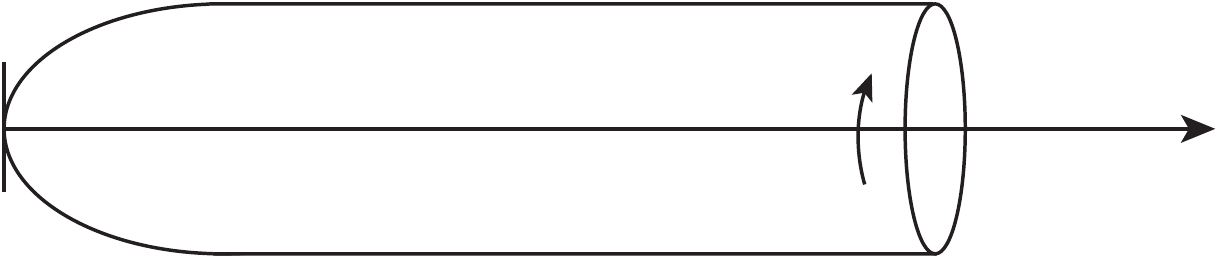}};
      \begin{scope}[shift={(-8,-2)},every node/.style={anchor=base west}]
          \draw (5,4) node[]{
            \begin{minipage}{10em}
              undeformed\\ \ac{tn}
            \end{minipage}}; \draw (5,0) node[]{\acs{ttrap}};

        \begin{small}
        
          \draw (13,2.2) node[]{\( r \)}; \draw (10.5,2.4)
          node[anchor={base east}]{\( \theta \)}; \draw (10.5,1.4)
          node[anchor={base east}]{\( \phi = \theta - 2 \wt R \epsilon
            \tilde u\)};
         
          \draw (1.5,3) node[]{\(\setR^4\)}; \draw (1,.5)
          node[]{fluxtrap};
        
          \draw (11,3.5) node[]{\(\setR^3 \times S^1\)}; \draw (11,0)
          node[]{
            \begin{minipage}{10em}
              \(\setR^3 \times S^1 \times S^1 \) plus \\ constant \( B \) field
            \end{minipage}
          };
        \end{small}
      \end{scope}
    \end{tikzpicture}
    \caption{The \ac{tn} interpolates between \( \setR^4 \) and \( \setR^3 \times S^1 \), while the corresponding \acs{ttrap} interpolates between the fluxtrap in flat space (tip of the cigar) and \( \setR^3 \times S^1 \) with a constant \( B \) field.}
  \label{fig:Taub-trap}
\end{figure}
Adding a \D4--brane wrapped on the \ac{tn} and bounded by two \NS5s in \( x^6 \) preserves the Killing spinors that satisfy
\begin{equation}
  \etaA^L = \Gamma_{\D4} \etaA^R \,, 
\end{equation}
where \( \Gamma_{\D4} \) is the pullback of \( \gamma_{m_1 \dots m_5} \) onto the \D4.
In other words, the four preserved supersymmetry generators are given by
\begin{equation}
  \eta_{\epsilon} = \etaA^L + \etaA^R = \left( \Id + \Gamma_{\D4} \right) \etaA^R = \left( \Id + \Gamma_{\D4} \right) \Gamma_u \etaB^R = \frac{\sqrt{V(r)} + \epsilon \gamma_{39}}{\sqrt{\epsilon^2 + V(r)}} \eta_{\epsilon = 0}\, .
\end{equation}
In this sense the \( \Omega  \)--deformation can be understood as a rotation in the \( (x^3,x^9) \)  plane by the angle
\begin{equation}
  \tan \frac{\vartheta}{2} = \frac{\epsilon}{\sqrt{ V(r)}}
\end{equation}
that acts on the spinors as
\begin{equation}
  \eta_{\epsilon} = \exp[ \frac{\vartheta}{2} \gamma_{39} ] \eta_{\epsilon = 0} \, .
\end{equation}
The \acs{ttrap} interpolates between two remarkable backgrounds:
\begin{itemize}
\item For \( r \to 0 \), the potential is \( V(r) \sim Q/r \) and the solution becomes
  \begin{subequations}
    \begin{align}
      \di s^2 &= \frac{Q}{r} \di \mathbf{r}^2 + \frac{r}{Q + \epsilon^2 r} \left( \di \phi + Q \cos \omega \di \psi \right)^2 + \frac{Q}{Q + \epsilon^2 r} (\di x^9)^2 + \di \mathbf{x}^2_{4 \dots 8}\,, \\
      B &= \frac{\epsilon r}{Q + \epsilon^2 r } \left( \di \phi + Q \cos \omega \di \psi \right) \wedge \di x^9 \,,\\
      \eu^{-\Phi} &= \sqrt{ 1 +
        \frac{\epsilon^2 r}{Q}}\,.
    \end{align}
  \end{subequations}
  After the changes of variables
  \begin{align}
    \begin{cases}
      r = Q \left( \rho_1^2 + \rho_2^2 \right)\,, \\
      \omega = 2 \arctan \frac{\rho_2}{\rho_1}\,, \\
      \phi = Q \left( \theta_2 - \theta_1 \right)\,, \\
      \psi = \theta_1 + \theta_2 \,,
    \end{cases} && \text{and} &&
    \begin{cases}
      x^0 + \im x^1 = \rho_1 \eu^{\im \theta_1}\,, \\
      x^6 + \im x^7 = \rho_2 \eu^{\im \theta_2} \,,
    \end{cases}
  \end{align}
  this is precisely the fluxtrap in flat space of Eq.~(\ref{eq:IIA-OneEpsilonTopologicalBulk}).
\item For \( r \to \infty \), the potential is \( V(r) \sim 1/\rTN^2 \) and the background is given by
  \begin{subequations}
    \label{eq:flat-space-constant-B}
    \begin{align}
      \di s^2 &= \frac{\di \mathbf{r}^2}{\rTN^2} + \frac{\rTN^2}{1 + \epsilon^2 \rTN^2} \di \phi^2 + \frac{(\di x^9)^2}{1 + \epsilon^2 \rTN^2} + \di \mathbf{x}^2_{4 \dots 8}\,, \\
      B &= \frac{\epsilon \rTN^2}{1 + \epsilon^2 \rTN^2} \di \phi \wedge \di x^9\,, \\
      \eu^{-\Phi} &=  \sqrt{ 1 +
        \epsilon^2 \rTN^2}\,.
    \end{align}
  \end{subequations}
  This is flat space (to be precise \( \setR^8 \times T^2 \)) with a constant \( B  \) field. Supersymmetry is restored, in the sense that in this limit there are thirty-two supercharges corresponding to the following Killing spinors:
  \begin{equation}
    \begin{cases}
      \etaA^L = \left( \Id + \Gamma_{11} \right) \eta_0 \, ,\\[.2em]
      \etaA^R = \displaystyle{\frac{\Id+ \epsilon \rTN \gamma_{39}}{\sqrt{1+\epsilon^2 \rTN^2}}} \left( \Id - \Gamma_{11} \right) \gamma_9 \eta_1 \, ,
    \end{cases}
  \end{equation}
  which means that the \( \Omega \)--deformation acts on the supersymmetry generators for the theory of a \D4--brane as a rotation:
  \begin{equation}
    \eta_{\epsilon} = \exp[ \frac{\vartheta}{2} \gamma_{39} ] \eta_{\epsilon = 0} \, ,
  \end{equation}
  where
  \begin{equation}
    \tan \frac{\vartheta}{2} = \epsilon \rTN \, .
  \end{equation}
\end{itemize}

The \ac{tn} geometry interpolates between \( \setR^4 \) for \( r \to 0 \) and \( \setR^3 \times S^1 \) for \( r \to \infty \) where the \( S^1 \) is a circle of radius \( \rTN \) described by \( \theta \). Adding a fluxtrap we obtain the \acs{ttrap} background that interpolates between the flat fluxtrap and flat space with a constant \( B  \) field. This is not surprising because in the large-\( r \) limit the identifications are done on a decoupled \( S^1 \) and the fluxtrap is realized by T--duality on a torus with shear (generated by \( (u, \phi) \)) which results in a constant Neveu--Schwarz field (see Figure~\ref{fig:Taub-trap}).
In this sense our construction relates the usual field-theory interpretation of the \( \Omega \)--deformation recovered in Section~\ref{sec:omega-def} (\( r \to 0 \)) and the alternative description of~\cite{Nekrasov:2010ka} (\( r \to \infty \)). In the latter limit, the T--dual background in Eq.~\eqref{eq:flat-space-constant-B} captures directly the deformations of metric and coupling constant and explains the «~\emph{non-trivial transformations of the observables}~».

T--duality on a torus with shear is also the first clue for the non-commutativity that we expect based on the observations in~\cite{Aganagic:2003qj} and~\cite{Nekrasov:2009rc}. In order to turn this clue into a precise statement, valid for finite values of \( \epsilon \) and for all values of \( r \), we need to pass to an equivalent string-theoretical description of the fluxtrap. As a first step we lift it to M--theory:
\begin{subequations}
  \begin{align}
      \di s^2 &=  \hspace{-.5em} \mbox{
          \begin{small}
            \(\displaystyle  \left(1 + \frac{\epsilon^2}{V(r)} \right)^{1/3}\) \end{small}}
      \hspace{-.2em}\left[ V(r) \di \mathbf{r}^2 + \hspace{-.4em}
        \mbox{
          \begin{small}
            \(\displaystyle\frac{\left(\di \phi + Q \cos \omega \di
                \psi \right)^2 + V(r) \left( (\di x^9)^2 + (\di
                x^{10})^2 \right)}{V(r) + \epsilon^2}\)
          \end{small}} \hspace{-.2em} + \di \mathbf{x}_{4
          \dots 8}^2 \right]\\
    C_3 &= \frac{\epsilon}{V(r) + \epsilon^2} \left( \di \phi + Q \cos
      \omega \di \psi \right) \wedge \di x^9 \wedge \di x^{10}\,.    
  \end{align}
\end{subequations}
This picture becomes particularly clear in the \( \epsilon \to 0 \) limit. The metric is \( \mathrm{TN}_Q \times \setR^7 \)
and the four--form flux is
\begin{equation}
  F_4 = \epsilon \, \wTN \wedge \di x^9 \wedge \di x^{10} \, ,
\end{equation}
where \( \wTN \) is the unique two-form on the \ac{tn} that is invariant under the triholomorphic \( U(1) \) isometry:
\begin{equation}
  \label{eq:omega-TN}
  \wTN = \di \left[ \frac{\di \phi + Q \cos \omega \di \psi}{V(r)} \right] \, .
\end{equation}
There are three natural circles that can be used to reduce the M--theory background to \tIIA. Reducing on \( x^{10} \) leads back to the fluxbrane on \ac{tn}. The same happens when reducing on \( x^9 \), as already observed in the previous section. The third alternative consists in reducing along \( \phi \). The resulting bulk contains a one-form and a three-form:
\begin{subequations}
  \begin{align}
    \begin{split}
      \di s^2 &=  V(r)^{1/2} \di \mathbf{r}^2 + V(r)^{-1/2} \left[ \di  \mathbf{x}^2_{4 \dots 8} + \frac{ (\di x^9)^2 + (\di x^{10})^2}{1 + \epsilon^2 \|U\|^2}  \right] = \\
      &=  V(r)^{1/2} \di \mathbf{r}^2 + V(r)^{-1/2} \left[ \di  \mathbf{x}^2_{4 \dots 8} + \frac{V(r) }{V(r) + \epsilon^2} \left( (\di x^9)^2 + (\di x^{10})^2\right) \right] \,,
    \end{split}\\
    B &= \frac{\epsilon}{V(r) \left( 1+  \epsilon^2 \|U\|^2 \right)} \di x^9 \wedge \di x^{10} =  \frac{\epsilon}{V(r) + \epsilon^2} \di x^9 \wedge \di x^{10} \,, \\
    \eu^{-\Phi} &= V(r)^{3/4} \sqrt{1+ \epsilon^2 \|U\|^2} =  V(r)^{1/4} \sqrt{V(r) + \epsilon^2} \,,\\
    A_1 &= Q \cos \omega \di \psi\,, \\
    A_3 &= B \wedge A_1\,.
  \end{align}
\end{subequations}
This background is the \( \Omega \)--deformation of the theory of \( Q \)  \D6--branes extended in \( (x^4, \ldots, x^{10}) \). 

An equivalent description is obtained by applying the \ac{sw} map~\cite{Seiberg:1999vs} to the \D6--brane theory in order to turn the \( B \)--field into a non-commutativity parameter:
\begin{equation}
  \left( \hat g + \hat B \right)^{-1} = \tilde g^{-1} + \Theta\,,
\end{equation}
where \( \hat g \) and \( \hat B \) are the pullbacks of metric and \( B \)--field on the brane and \( \tilde g \) is the new effective metric for a non-commutative space satisfying
\begin{equation}
  [ x^i, x^j ] = \im \Theta^{ij} \, .  
\end{equation}
Applying this map to our case we find that the \( \epsilon \)--dependence of the metric is completely dropped and the \( B \)--field is turned into a non-commutativity between \( x^9 \) and \( x^{10} \):
\begin{align}
  \tilde g_{ij} \di x^i \di x^j &= V(r)^{-1/2} \di \mathbf{x}^2_{4 \dots 10}\,,\\
  [ x^9, x^{10} ] &= \im \epsilon\,.
\end{align}
Maybe surprisingly, all dependence on $\epsilon$ disappears from the \D6--brane theory and is turned into a constant non-commutativity parameter\footnote{Non-commutative gauge theories in Melvin backgrounds have already been discussed in~\cite{Hashimoto:2005hy}, albeit with a different brane configuration.}.We would like to stress that this is an exact result, valid for any finite value of $\epsilon$ and \( r \).

\bigskip

At this point, it is interesting to follow the fate of the branes whose dynamics reproduce the $\Omega$--deformed gauge theory. Start from the configuration of \D4/\NS5s given in Table~\ref{tab:D4-embedding-two-epsilon}, where the \ac{tn} space is extended in $(x^0,\dots,x^3)$. We have seen in the previous section that in the M--theory lift this configuration turns into a single \M5--brane extended in the directions $(x^0,\dots,x^3)$ and wrapped on a Riemann surface $\Sigma$ embedded in the $(s,v)$ plane. Reduction on $\phi$ turns the \M5--brane into a \D4--brane extended in~$\mathbf{r}$ and wrapped on \( \Sigma \), which is now embedded in the worldvolume of the \D6--brane. This is strictly true in the $\epsilon=0$ limit. Our findings above point towards the fact that for finite $\epsilon$ this picture remains the same, but this time the Riemann surface $\Sigma$ is embedded in a non-commutative complex plane where
\begin{equation}
  \boxed{%
  [s,v] = \im \epsilon\,.%
}
\end{equation}

As observed in Section~\ref{sec:topol-string-limit}, this background is related to topological strings. In this sense, our picture provides a geometric explanation for the fact that in this context the Riemann surface behaves «~\emph{as a subspace of a quantum mechanical \( s,v \) phase space}~»~\cite{Aganagic:2003qj}. Our setup should be contrasted with the one in~\cite{Dijkgraaf:2007sw}, where a similar explanation was offered for the non-commutativity. In this case the authors start from M--theory on \( \mathrm{TN} \times \wt X \times S^1 \), with Melvin identifications in the \ac{tn} and apply a sequence of dualities leading eventually to \M5--branes wrapped on a Riemann surface in \( \mathrm{TN} \times \setC^2 \times \setR^2 \times S^1 \) (\( \wt X \) is the mirror of the Calabi--Yau defined by \( x y + F(s,v) = 0 \), where \( F(s,v) = 0 \) is the Riemann surface \( \Sigma \)). 
As already stressed in Section~\ref{sec:topstring}, even though the geometry is the same as ours for \( \epsilon \to 0 \), there are important differences.
In particular the Melvin construction is realized on a different circle so that in~\cite{Dijkgraaf:2007sw} the \( F_4 \) flux has no~\( (1,1) \) components on the \( \setC^2 \) where \( \Sigma \) is embedded and its pullback on the \M5--brane vanishes.

\subsection{The NS limit and gauge/Bethe correspondence}
\label{sec:non-commutative-ns-limit}

In this section, we want to study a different fluxtrap background on the same $\mathrm{TN}_Q\times S^1\times \setR^5$ space with only a single $\epsilon$ on the \ac{tn} part. This corresponds to taking the \ac{ns}~limit $\epsilon_1=0$ on the \D4--brane gauge theory as discussed in Section~\ref{sec:NS}.

In order to impose the Melvin identifications we need to choose a different coordinate system for the \ac{tn} space in which its nature as a complex two-dimensional manifold is manifest (see Appendix~\ref{sec:taub-nut-coordinates} for details):
\begin{multline}
  \label{eq:TNmetric-complex}
  \di s^2 = \frac{V(\rho)}{Q} \left[ \rho_1^2 \rho_2^2 \left( \di \theta_1 + \di \theta_2 \right)^2 + \left( \rho_1^2 + \rho_2^2 \right) \left( \di \rho_1^2 + \di \rho_2^2 \right) \right] + \frac{Q}{V(\rho)} \left[ \frac{\rho_1^2 \di \theta_1 - \rho_2^2 \di \theta_2}{\rho_1^2 + \rho_2^2} \right]^2 \\
+ \di \rho_3^2 + \rho_3^2 \di\theta_3^2 + (\di \tilde x^9)^2 + \di \mathbf{x}_{6,7,8}^2  \,, 
\end{multline}
where
\begin{equation}
  V(\rho) = \frac{1}{\rTN^2} + \frac{Q}{\rho_1^2 + \rho_2^2} \, .
\end{equation}
We impose the identifications
\begin{align}
  \begin{cases}
    \tilde u \sim \tilde u + 2\pi n_u\,,\\
    \theta_1 \sim \theta_1+\epsilon_1 2\pi \wt R n_u \,,\\
    \theta_3 \sim  \theta_3+\epsilon_3 2\pi \wt R n_u \,,
  \end{cases} && n_u \in \setZ
\end{align}
where $\epsilon_1 = -\epsilon_3 = \epsilon$ in order to preserve supersymmetry. We introduce the disentangled variables
\begin{equation}
  \begin{cases}
    \phi_1 = \theta_1 - \epsilon_1 \wt R \tilde u\,,\\
    \phi_2 = \theta_2 \,,\\
    \phi_3 = \theta_3 - \epsilon_3 \wt R \tilde u\,.
  \end{cases}
\end{equation}
The corresponding rotational isometry generator is given by
\begin{subequations}
  \begin{align}
    U^i\del_i &= \del_{\phi_2} - \del_{\phi_3}\,,\\
    U_i\di x^i  &= \rho_2^2 \left[ \frac{\rho_1^2 V(\rho) \left( \di \phi_1 + \di \phi_2 \right)}{Q} + \frac{Q \left( - \rho_1^2 \di \phi_1 + \rho_2^2 \di \phi_2 \right)}{\left( \rho_1^2 + \rho_2^2 \right)^2 V(\rho)} \right] - \rho_3^2 \di\phi_3 \,,\\
    \|U\|^2 &= \rho_2^2 \left[ \frac{ Q \rho_2^2 }{\left(\rho_1^2 +
          \rho_2^2\right)^2 V(\rho) } + \frac{\rho_1^2 V(\rho)}{Q}  \right] + \rho_3^2\,.
  \end{align}
\end{subequations}
In the \( \rho \to 0 \) limit we find
\begin{equation}
  \|U\|^2 \to \rho_2^2 + \rho_3^2 \, ,
\end{equation}
consistently with the fact that the \ac{tn} space is asymptotically \( \setR^4 \) and we are back to the fluxtrap solution in Section~\ref{sec:NS}.

Now the chosen isometry acts as a linear rotation of a noncompact space, the fluxtrap breaks some of the supersymmetry and only preserves eight supercharges\footnote{The same eight Killing spinors are preserved under the more general identifications
\begin{align*}
  \begin{cases}
    \tilde u \sim \tilde u + 2\pi n_u\,,\\
    \theta_1 \sim \theta_1 + 2\pi \epsilon_1 \wt R n_u \,,\\
    \theta_2 \sim \theta_1 + 2\pi \epsilon_2 \wt R n_u \,,\\
    \theta_3 \sim \theta_3 + 2\pi \epsilon_3 \wt R n_u \,,
  \end{cases} && n_u \in \setZ
\end{align*}
with the condition \( \epsilon_1 + \epsilon_2 + \epsilon_3 = 0 \).
}.  This can be verified directly by starting from the expression of the 16 Killing spinors prior to the identifications: 
\begin{equation}
  \etaB = \wt \Gamma(\tfrac{\rho_1}{\rho_2}) \gamma_3 \exp[ \frac{\theta_1}{2} \gamma_{01} ] \exp[ \frac{\theta_2}{2} \gamma_{23} ] \exp[ \frac{\theta_3}{2} \gamma_{56} ] \left( \gamma_{01} + \gamma_{23} \right) \etaw \, ,
\end{equation}
where
\begin{equation}
  \wt \Gamma( \tfrac{\rho_1}{\rho_2} ) =  \sqrt{\frac{\rho_2}{\sqrt{\rho_1^2 + \rho_2^2}} + 1} \left(\frac{\rho_1}{\rho_2+\sqrt{\rho_1^2+\rho_2^2}} \gamma_0 - \gamma_2 \right) \, .  
\end{equation}
Introducing the disentangled variables \( \phi_k \) we can isolate the terms that depend explicitly on \( \tilde u \) and do not satisfy the Melvin boundary conditions: 
\begin{equation}
  \etaB = \wt \Gamma(\tfrac{\rho_1}{\rho_2}) \gamma_3 \exp[ \frac{\phi_1}{2} \gamma_{01} ] \exp[ \frac{\phi_2}{2} \gamma_{23} ] \exp[ \frac{\phi_3}{2} \gamma_{45} ] \exp[ \frac{\epsilon \wt R \tilde u}{2} \left( \gamma_{01} - \gamma_{45} \right)] \left( \gamma_{01} + \gamma_{23} \right) \etaw \,.
\end{equation}
The \( \tilde u \)-dependent part is projected out via \( (\gamma_{01} + \gamma_{45} )\). The final result is that \( 1/2 \) of the supersymmetries are broken and after T--duality the eight Killing spinors are
\begin{equation}
  \begin{cases}
    \etaA^L = \left( \Id + \Gamma_{11} \right)  \Gamma(\tfrac{\rho_1}{\rho_2}) \gamma_3 \exp[ \frac{\phi_1}{2} \gamma_{01} ] \exp[ \frac{\phi_2}{2} \gamma_{23} ] \exp[ \frac{\phi_3}{2} \gamma_{56} ] \left( \gamma_{01} + \gamma_{45} \right) \left( \gamma_{01} + \gamma_{23} \right) \eta_0 \, ,\\
    \etaA^R = \left( \Id - \Gamma_{11} \right) \Gamma_u \Gamma(\tfrac{\rho_1}{\rho_2}) \gamma_3 \exp[ \frac{\phi_1}{2} \gamma_{01} ] \exp[ \frac{\phi_2}{2} \gamma_{23} ] \exp[ \frac{\phi_3}{2} \gamma_{56} ] \left( \gamma_{01} + \gamma_{45} \right) \left( \gamma_{01} + \gamma_{23} \right) \eta_1 \, ,
  \end{cases}
\end{equation}
where \( \Gamma_u \) is the gamma matrix in the direction of the T--duality.

At this point we can proceed as in the previous section. Write down the \tIIA fields after T--duality and the corresponding M--theory lift which we reduce, this time on \( \phi_2 \). The final result after the flip can be expressed in terms of two functions:
\begin{align}
  f_1 (\rho_3) &= \sqrt{1+\epsilon^2 \rho_3^2}\,, \\
  f_2 (\rho_1, \rho_2 ) &= \sqrt{V(\rho)}\sqrt{\|U\|^2 - \rho_3^2}\, .%
\end{align}
The Neveu--Schwarz sector is given by
\begin{subequations}
  \begin{align}
    \begin{split}
       \di s^2 &= f_1(\rho_3) f_2(\rho_1, \rho_2) \Bigg[
      4 V(\rho)^{1/2} \left(\left( \rho_1^2 + \rho_2^2 \right) \left(
          \di \rho_1^2 + \di \rho_2^2 \right) + \frac{4 Q^2\rho_2^2
          \rho_1^2 \di \phi_1^2}{f_2(\rho_1,\rho_2)^2}\right) \\ 
      & \hspace{9em} +  V(\rho)^{-1/2} \left( \di \rho_3^2 + \di \mathbf{x}^2_{6,7,8} +
        \frac{\rho_3^2 \di \phi_3^2 }{f_1(\rho_3)^2 } + \frac{( \di
          x^9)^2 + (\di x^{10})^2 }{1+ \epsilon^2 \|U\|^2}
      \right) \Bigg] \,,
    \end{split}
    \\
    B &= \epsilon \frac{f_2^2(\rho_1, \rho_2) }{V(\rho) \left(1+\epsilon^2 \|U\|^2 \right)} \di x^9 \wedge \di x^{10} \, , \\
    \eu^{-\Phi} &= \frac{V(\rho)^{3/4} \sqrt{ 1 + \epsilon^2 \|U\|^2}}{f_1(\rho_3)^{3/2} f_2(\rho_1,\rho_2)^{3/2}} \, .
  \end{align}
\end{subequations}
The structure of the solution is precisely the same as in the previous case, and we can recognize the terms corresponding to \( Q \)  \D6--branes extended in \( (\rho_3, \phi_3, x^6, \dots, x^{10}) \). We can now apply the \ac{sw} map. Once more the \( B \)--field is traded for a constant non-commutativity parameter between \( x^9  \) and \( x^{10} \):
\begin{equation}
  \hat g_{ij} \di x^i \di x^j =\frac{ f_2(\rho_1,\rho_2)}{V(\rho)^{1/2}} \left[ f_1(\rho_3) \left(\di \rho_3^2 + \di \mathbf{x}^2_{6,7,8} \right) + \frac{\rho_3^2 \di \phi_3^2 + ( \di x^9)^2 + (\di x^{10})^2}{f_1(\rho_3)}\right]\,,
\end{equation}
\begin{equation}
  [ x^9, x^{10} ] = \im \epsilon \,.
\end{equation}
Even though the in this case the dependence on \( \epsilon \) is not completely dropped from the metric (which is flat around \( \rho \to 0 \), where the partition functions are evaluated), remarkably we still find that the non-commutativity parameter is constant and equal to \( \epsilon \), without need for approximations.

\bigskip

The study of the brane dynamics in this background follows closely the discussion in the previous section. Following the dualities, the \D4/\NS5 system that corresponds to the gauge theory in the \ac{ns} limit described in Section~\ref{sec:NS} is lifted to a single \M5--brane wrapping a Riemann surface \( \Sigma \) and is then reduced to a \D4--brane wrapped on a Riemann surface embedded on the \D6--brane. After the \ac{sw} map, this can be understood as a Riemann surface on a non-commutative two-dimensional manifold satisfying
\begin{equation}
  [ s, v ] = \im \epsilon \, .  
\end{equation}
This provides a geometrical interpretation for the fact that \( \Omega \)--deformed four-di\-men\-sion\-al gauge theories in the \ac{ns} limit are associated to quantum integrable models with~\( \hbar = \epsilon \)~\cite{Nekrasov:2009rc,Dijkgraaf:2008ua}.

\section{Conclusions}\label{sec:conc}

In this article, we have presented a string theory realization of the $\Omega$--deformation of gauge theory. Our framework has the virtue of capturing general $\epsilon$--deformations which include the various special cases discussed in the literature such as the topological string and the \acl{ns} limit. It provides moreover a \emph{geometric} interpretation for the properties of $\Omega$--deformed gauge theories. Given the stringy nature of the construction, the methods of string theory are applicable, which in this case are often more powerful and transparent than their gauge theory equivalents.

T--duality plays a key role in our construction by making the effects of the underlying Melvin background evident.
It is possible to lift the fluxtrap background to M--theory, relating it thus to the famous and elusive $(2,0)$ gauge theory in six dimensions. The M--theory lift is also instrumental for the 9--11 flip via which we can connect the fluxtrap background to non-commutative gauge theory. We find in particular that the Riemann surface $\Sigma$ on which the M5--brane is wrapped is now embedded in a non-commutative complex plane with non-commutativity parameter \( \epsilon = \hbar \), matching up neatly with the \emph{quantum spectral curve} of the integrable system discussed in~\cite{Nekrasov:2009rc} which also plays a prominent role in topological string theory. Again, the fluxtrap construction gives a geometrical interpretation also to the \emph{quantum integrable system}.  

A question that can maybe be attacked from here is whether our string picture can be used to shed some light also on the \textsc{agt} conjecture~\cite{Alday:2009aq}.  In one respect this
is surely the case, in that our realization gives an algorithmic construction
of the modified couplings realizing the general $\Omega$--deformation of an
$\mathcal{N} = 2$ theory.

In principle these couplings can be inferred on general
grounds through
considerations of self-consistency, by requiring the preservation of certain supersymmetries that of gauge theory dynamics after
the deformation, a method used for instance in~\cite{Hama:2010av, Hama:2011ea} to derive the modified geometry of $\Omega$--deformed three-dimensional theories and defect theories.
The application of this abstract method to four-dimensional theories has not yet produced a derivation for the deformed geometry and twisted couplings corresponding to the $\Omega$--deformation
of general four-dimensional $\mathcal{N}=2$ gauge theories with general $\e$--parameters.

The string solutions written down here allow the deformed couplings to be read off from the branes' coupling to the modified spacetime metric and
other supergravity background fields in \eqref{eq:general-fluxtrap},
giving a physical realization to the method of  \cite{Festuccia:2011ws}.
 For the gauge and adjoint degrees of freedom, our solution 
gives a straightforward prescription for the $\Omega$--deformation of
the action, through the Born--Infeld and Chern--Simons action of the \D4--branes on
which the gauge theory degrees of freedom propagate.  
In particular, for the refined case $\e\lll 2 \neq - \e\lll 1$
the \ac{dbi} action generates explicit terms of order higher than $|\e|\sqd$ that would be at best cumbersome to  deduce abstractly by demanding the preservation of a conserved twisted supercharge.  The deformation of the action of the fundamental
and bifundamental degrees of freedom of the \NS5/\D4 system, while not manifest in
the \ac{dbi} action, is determined by open string
worldsheet physics in the fluxtrap background; a useful direction would be to learn to extract
those deformed couplings to the deformed closed string fields in an efficient manner. 

\subsection*{Acknowledgements}
 
It is our pleasure to thank Lotte Hollands, Can Kozçaz, Neil Lambert and Wolfgang Lerche for useful discussions, and
Kazuyuki Furuuchi for correspondence.

The work of S.H. was supported by the World Premier International Research Center Initiative, \textsc{mext}, Japan, and also by a Grant-in-Aid for
Scientific Research (22740153) from the Japan Society for Promotion of Science (\textsc{jsps}).  S.H.
is also grateful to the theory group at \textsc{cern} for hospitality while this work was in progress.

D.O. and S.R. would like to thank \textsc{ipmu} for hospitality during the final stages of this work, as well as the organizers of the workshop ``\emph{New perspectives on supersymmetric gauge theories}'' (Munich, February 2012) and the organizers of the workshop ``\emph{Problemi Attuali di Fisica Teorica}'' (Vietri sul mare, March 2012) where some of the results of this paper were announced.

\clearpage
\appendix

\section{Supersymmetry conventions}
\label{sec:supersymm-conv}

The condition for preserving supersymmetry in eleven-dimensional supergravity is the vanishing of the variation of the gravitino \( \Psi \):
\begin{equation}
  \delta \Psi_m = \left[\nabla_m + \frac{1}{288} \left( \Gamma\indices{_m^{m_1}^\dots^{m_4}} - 8 \delta\indices{_m^{m_1}} \Gamma^{m_2 m_3 m_4} \right) F_{m_1 \dots m_4}  \right] \etaM = 0 \, ,
\end{equation}
where \( F_4 = \di A_3 \) is the flux of the three-form field, and the covariant derivative acts on spinors as \(  \nabla_m \eta = \del_m \eta + \frac{1}{4} \omega\indices{_m^a^b} \gamma_{ab} \).

In order to reduce on \( x^{10} \) we write the metric as
\begin{equation}
  \textstyle \di s^2_{11} = \eu^{-2 \Phi/3} \di s^2_{10} + \eu^{4 \Phi/3} \left(\di x^{10} + A_1 \right)^2,
\end{equation}
and the three-form field as
\begin{equation}
    A_3 = C_3 + B \wedge \di x^{10} \, .
\end{equation}
The vielbein \( \EM \) is written in terms of the ten-dimensional \( \EA \) as
\begin{equation}
  \begin{cases}
    \EM^a = \eu^{-\Phi/3} \EA^a & \text{for \( a = 0, \dots, 9 \);} \\
    \EM^{10} = \eu^{2\Phi/3} \left( \di x^{10} + A_1 \right) \, ,
  \end{cases}
\end{equation}
and the gravitino is decomposed into a dilatino \( \rTN \) and a ten-dimensional gravitino \( \psi \):
\begin{equation}
  \begin{cases}
    \Psi_{10} = \frac{1}{3} \eu^{\Phi/6} \Gamma_{11} \rTN \, , \\
    \Psi_{m} = \eu^{\Phi/6} \left( \psi_m - \frac{1}{6} \Gamma_m
      \rTN \right) \, .
  \end{cases}
\end{equation}
Then the variation \( \delta \Psi_m \) becomes:
\begin{equation}
  \begin{cases}
    \delta \rTN = \left[ \slashed{\del} \Phi - \frac{1}{12} \slashed{H} \Gamma_{11} - \frac{1}{8} \eu^{\Phi} \left( 3 \slashed{F}_2 \Gamma_{11} - \frac{1}{12} \slashed{G} \right) \right] \etaA \, , \\
    \delta \psi_m = \left[ \nabla_m - \frac{1}{8} H_{m m_1 m_2} \Gamma^{m_1 m_2} \Gamma_{11} - \frac{1}{8} \eu^{\Phi} \left(\frac{1}{2} \slashed{F}_2 \Gamma_m \Gamma_{11} - \frac{1}{4!} \slashed{G} \Gamma_m \right)\right] \etaA \, ,
  \end{cases}
\end{equation}
where \( H = \di B \), \( G = \di A_3 - H \wedge A_1 \) and the ten-dimensional Killing spinor is related to the eleven-dimensional one by:
\begin{equation}
  \etaA = \eu^{\Phi/6} \etaM \, .
\end{equation}

T--duality in the direction \( u \) turns the \tIIA background into a \tIIB one. In absence of Ramond--Ramond fields the variation of the \tIIB dilatino and gravitino take the same form as in \tIIA. If we choose the \tIIA Vielbein as
\begin{equation}
  \begin{cases}
    \EB\indices{^a_{\tilde u}} = \frac{\alpha'}{g_{uu}} \EA\indices{^a_u} \, ,\\
    \EB\indices{^a_{\sigma}} = \EA\indices{^a_\sigma} - \frac{g_{\sigma u} + B_{\sigma u}}{g_{uu}} \EA\indices{^m_u}& \text{for \( x^\sigma \neq  u \),} 
  \end{cases}
\end{equation}
if \( \etaA \) does not depend on \( u \), the \tIIB Killing spinor is~\cite{Hellerman:2011mv,Reffert:2011dp}:
\begin{equation}
  \etaB = \left[ \left( \Id + \Gamma_{11} \right) + \im  \Gamma_u \left( \Id - \Gamma_{11} \right)\right] \etaA \, ,  
\end{equation}
where \( \Gamma_u \) is the gamma matrix in the direction \( u \) normalized to one.

\section{Taub--NUT spaces}
\label{sec:taub-nut-coordinates}

\paragraph{Coordinate systems.}

We use two coordinate systems for the \ac{tn} space. In the \ac{gh} system the space is seen as a singular circle fibration over \( \setR^3 \) and in the second (cylindrical) system, the space is seen as a two-dimensional complex manifold. They are, respectively
\begin{align}
  \di s^2 &= V(r) \left( \di r^2 + r^2 \di \omega^2 + r^2 \sin \omega \di \psi \right) + \frac{1}{V(r)} \left( \di \theta + Q \cos \omega \di \psi \right)^2 \, ,& V(r) &= \frac{1}{\rTN^2} + \frac{Q}{r}
\end{align}
and
\begin{align}
  \label{eq:TNmetric-complex}
  \di s^2 &= \frac{V( u )}{4Q} \di \mathbf{u} \cdot \di \mathbf{u} + \frac{Q}{4V( u)} \left(\frac{\Im ( \bar z_1 \di z_1 - \bar z_2 \di z_2) }{\abs{z_1}^2 + \abs{z_2}^2 } \right)^2  \,, & V(u) &= \frac{1}{\rTN^2} + \frac{Q}{\sqrt{u_1^2 + u_2^2 + u_3^2}}
\end{align}
where
\begin{align}
  u_1 &= 2 \Re (z_1 z_2)\,, & u_2 &= 2 \Im (z_1 z_2)\,, & u_3 &= \abs{z_1}^2 - \abs{z_2}^2 \, .
\end{align}
In order to find the transformations between the two metrics, it is convenient to introduce polar coordinates on the complex plane
\begin{align}
  z_1 &= \rho_1 \eu^{\im \theta_1}\,, & z_2 &= \rho_2 \eu^{\im \theta_2} \,,
\end{align}
then the metric in Eq.~(\ref{eq:TNmetric-complex}) becomes
\begin{equation}
  \di s^2 = \frac{V(\rho)}{Q} \left( \rho_1^2 \rho_2^2 \left( \di \theta_1 + \di \theta_2 \right)^2 + \left( \rho_1^2 + \rho_2^2 \right) \left( \di \rho_1^2 + \di \rho_2^2 \right) \right) + \frac{Q}{V(\rho)} \left[ \frac{\rho_1^2 \di \theta_1 - \rho_2^2 \di \theta_2}{\rho_1^2 + \rho_2^2} \right]^2
\end{equation}
and
\begin{equation}
  V (\rho) = \frac{1}{\rTN^2 } + \frac{Q}{\rho_1^2 + \rho_2^2}\,.
\end{equation}
The coordinates are changed according to:
\begin{align}
  \begin{cases}
    \rho_1 = \sqrt{r} \cos \frac{\omega}{2} \\
    \rho_2 = \sqrt{r} \sin \frac{\omega}{2} \\
    \theta_1 = \frac{\psi + \theta}{2} \\
    \theta_2 = \frac{\psi - \theta}{2}     
  \end{cases} && 
  \begin{cases}
    r = \rho_1^2 + \rho_2^2 \\
    \omega = 2 \arctan \frac{\rho_2}{\rho_1} \\
    \theta = \theta_1 - \theta_2 \\
    \psi = \theta_1 + \theta_2 \,.   
  \end{cases}
\end{align}
From this explicit form it is clear that Melvin identifications in \( \theta_1  \) and \( \theta_2 \) with coefficient \( \epsilon_1 = - \epsilon_2 = \epsilon \) are equivalent to a single Melvin identification in \( \theta \) with coefficient \( \epsilon \).

\bigskip

The (near-horizon) limit \( r \to 0 \) is transparent in the cylindrical coordinate system. We find that \( V(\rho) \sim Q \left(\rho_1^2 + \rho_2^2 \right)^{-1} \) and the metric becomes the flat metric in cylindrical coordinates, \begin{equation}
  \di s^2  \sim  \di \rho_1^2 + \di \rho_2^2 + \rho_1^2 \di \theta_1^2 + \rho_2^2 \di \theta_2^2 \, .
\end{equation}
The large \( r \) limit \( r \to \infty \) is more clear in the \ac{gh} coordinates, where we have \( V (r) \sim \rTN^{-2} \) and the metric is asymptotically the cartesian product of \( \setR^3 \) with a circle or radius \( \rTN \).

\paragraph{Supersymmetry.}

Consider a \ac{tn} metric in \ac{gh} coordinates. Choose the vielbein
\begin{equation}
  \begin{aligned}
    \EB^0 &= \sqrt{V(r)} \di r \, ,&  \EB^1 &= r \sqrt{V(r)} \di \omega \, ,\\
    \EB^2 &= r \sin \omega\sqrt{V(r)}\di\psi \, ,& \EB^3 &= \frac{1}{\sqrt{V(r)}}\left( \di \theta + Q \cos \omega \di \psi\right) \, .
  \end{aligned}
\end{equation}
The Killing spinors solve the equation
\begin{equation}
  \del_m \eta + \frac{1}{4} \omega\indices{_m^a^b} \gamma_{ab} \eta = 0   \,,
\end{equation}
and take the form
\begin{equation}
  \label{eq:TN-spinors-GHcoord}
  \eta = \exp[\frac{\omega}{2} \gamma_{01}] \exp[ \frac{\psi}{2} \gamma_{23} ] \left( \gamma_{01} + \gamma_{23} \right) \etaw \, ,
\end{equation}
where $\etaw$ is constant Weyl spinor. Note the projector $\left(\Id-\gamma_{0123}\right)$ and the fact that $\eta$ does not depend on the fiber direction $\theta$.

\bigskip

In cylindrical coordinates a possible vielbein is:
\begin{equation}
  \begin{aligned}
    \EB^0 &= \sqrt{\left( \rho_1^2 + \rho_2^2 \right) \frac{V(\rho)}{Q}} \di \rho_1\, , &
    \EB^1 &= \rho_1 \rho_2 \sqrt{\frac{V(\rho)}{Q}} \left(\di \theta_1 + \di\theta_2 \right)  \, ,\\
    \EB^2 &= \sqrt{\left( \rho_1^2 + \rho_2^2 \right) \frac{V(\rho)}{Q}} \di \rho_2\, ,&
    \EB^3 &= \frac{\sqrt{Q}\left(\rho_1^2\di\theta_1 - \rho_2^2\di\theta_2 \right)}{2\left( \rho_1^2+\rho_2^2 \right)\sqrt{V(\rho)}} \,.
  \end{aligned}
\end{equation}
The corresponding Killing spinors read:
\begin{equation}
  \eta = \sqrt{\frac{\rho_2}{\sqrt{\rho_1^2 + \rho_2^2}} + 1} \left(\frac{\rho_1}{\rho_2+\sqrt{\rho_1^2+\rho_2^2}} \gamma_0 - \gamma_2 \right) \gamma_3 \exp [ \frac{\theta_1}{2}\gamma_{01}] \exp [ \frac{\theta_2}{2} \gamma_{23}] \left( \gamma_{01} + \gamma_{23} \right) \etaw \, .
\end{equation}
Note that $\eta$ only depends on $\theta_1+\theta_2$ and $\rho_2/\rho_1$. The projector remains the same. This is not surprising given that the change of coordinates is $\theta_1+\theta_2=\psi$ and $\rho_2/\rho_1=\tan (\omega/2)$.

\section{General fluxtrap action for complex epsilon}
\label{sec:compeps}

In this note, we have concentrated on $\Omega$--deformations involving real $\epsilon$--parameters. The construction with complex $\epsilon$ is similar to the one in Sec.~\ref{sec:fluxtrap}, with the difference that we now have to perform two T--dualities (in the $x^8$ and $x^9$ directions) with two associated sets of identifications in the same directions $\theta_k$, but with two independent deformation parameters $m_{8,k}$ and $m_{9,k}$ which combine to form the now complex deformation parameter $\epsilon_k$. For the details, we refer the reader to~\cite{Reffert:2011dp}.

As in Sec.~\ref{sec:generaleps}, the general action can now be written down. 
The complex counterpart of the action in Eq.~(\ref{eq:Two-epsilon-gauge}) is given by
\begin{multline}
  \mathscr{L}_{\epsilon_1, \epsilon_2} =   \frac{1}{4 g_4^2}\Big( 1 + \| F \|^2 + \frac{1}{2} \| \di \varphi +  \epsilon \imath_U F + \epsilon \abs{\epsilon}^2 (\imath_U \imath_{\bar U} F) U \|^2 \\
 + \frac{\abs{\epsilon}^2}{8} \| \imath_{\bar U} \di \varphi - \imath_U \di \bar \varphi \|^2  + \frac{\abs{\epsilon}^2}{2} \left( \imath_U \imath_{\bar U} F \right)^2  \left( 3 + \| \epsilon\,U \|^2 \right) \Big),
\end{multline}
or, in components (double indices are summed over):
\begin{multline}
  \mathscr{L}_{\epsilon_1, \epsilon_2 } = \frac{1}{4 g_4^2} \Big[ 1 + F_{ij} F^{ij} \\ + \frac{1}{2} \left( \partial_i \varphi +  \epsilon\,U^k F_{ki} +\epsilon \abs{\epsilon}^2 U^k \bar U^l F_{kl} U_i \right) \delta^{ij} \left( \partial_j \bar \varphi +  \bar \epsilon\,\bar U^k F_{kj} - \bar \epsilon \abs{\epsilon}^2 U^k \bar U^l F_{kl} \bar U_j \right) \\ 
  - \frac{1}{8} \left( \bar\epsilon\, \bar U^i \partial_i \varphi - \,\epsilon\,U^i \partial_i \bar \varphi \right)^2 + \frac{\abs{\epsilon}^4 }{2} \left(  U^k \bar U^l F_{kl} \right)^2 \left( 3 + \abs{\epsilon}^2 U^i \bar U_i \right) \Big] .
\end{multline}
Here, $U$ is the pullback of the Killing vector as in Eq.~(\ref{eq:pullU}), where we have in an abuse of notation used $U$ for the pullback instead of $\hat U$.

These actions contain more terms than the ones in Sec.~\ref{sec:generaleps}.
In the special case of \( \epsilon_1 / \epsilon_2 \in \mathbb{R} \), corresponding to \( \imath_U \imath_{\bar U} F \equiv 0 \) however, the action simplifies and is formally the same that we had found for a real \( \epsilon \) (which is the same as in~\cite{Nekrasov:2009rc}).  

It should be stressed that all expressions given in the main text of this article are formally correct in the case of complex $\epsilon$ with \( \imath_{U} \imath_{\bar U} = 0 \).

\section{Omega deformation of $\mathcal{N}=4$ SYM}
\label{sec:hypers}

Another straightforward generalization of our construction is obtained by removing the \NS5--branes from the setup in Section~\ref{sec:omega-def} and compactifying the theory in the direction \( x^6 \). The effective description for the dynamics of the resulting \D3--branes is the $\Omega$--deformed Lagrangian of \( \mathcal{N} = 4 \) super Yang--Mills:
\begin{multline}
  \mathcal{L}=\frac{1}{4g^2}\Big[ F_{ij}F^{ij} + \frac{1}{2} \left( \del^i \varphi + V^k F\indices{_{k}^{i}} + V^k\bar V^l F_{kl} V^i\right) \left( \del_i \bar \varphi + \bar V^kF_{ki} + \bar V^k V^l F_{kl} \bar V_i \right) + \\
 - \frac{1}{8}( \bar V^i \del_i \varphi - V^i \del_i \bar \varphi)^2 + \frac{1}{2}( V^k \bar V^lF_{kl})^2 \left( 3 + V^k \bar V_k \right) + \\
 + \frac{1}{4} \left( \delta^{ij} + V^i \bar V^j \right) \left( \del_i z \del_j \bar z + \text{c.c.} \right) + \frac{1}{4} \left( \delta^{ij} + V^i \bar V^j \right) \left( \del_i w \del_j \bar w + \text{c.c.} \right) + \\
 + \frac{1}{2 \im}  \left( \epsilon_3 \bar V^i + \bar \epsilon_3 V^i \right)  \left( \bar w \del_i  w - \text{c.c.} \right) + \frac{1}{2} \abs{\epsilon_3}^2 w \bar w \Big] \, ,
\end{multline}
where \( V = \epsilon \hat U = \epsilon_1 \left( \xi^0 \del_1 - \xi^1 \del_0 \right) + \epsilon_2 \left( \xi^2 \del_3 - \xi^3 \del_2 \right) \) and the fields \( w \) and \( z \) describe the oscillations of the \D3--brane respectively in \( x^4 + \im x^5 \) and \( x^6 + \im x^7 \) (see Table~\ref{tab:D3-embedding-two-epsilon}). The effect of the deformation on these two fields consists in a modification of the kinetic term. Moreover, the field \( w \) acquires a mass term (much like the twisted mass terms in~\cite{Hellerman:2011mv,Reffert:2011dp}) and a one-derivative term, which is allowed by the broken Poincaré invariance. The action and its properties deserve further study, but this goes beyond the scope of the present work.

\begin{table}
  \centering
  \begin{tabular}{lcccccccccc}
    \toprule
    \( x \)   & 0               & 1               & 2               & 3  & 4      & 5      & 6 & 7 & 8 & 9 \\
    \midrule
    fluxbrane & \ep{\epsilon_1} & \ep{\epsilon_2} & \ep{\epsilon_3} & \X & \X     & \X     & \T            \\
    \D3       & \X              & \X              & \X              & \X &        &        &   &   &   &   \\
    \midrule
    \( \xi \) & 0               & 1               & 2               & 3  & \ep{w} & \ep{z} & \ep{\varphi}  \\
    \bottomrule
  \end{tabular}
  \caption{\D3 brane realizing the \( \Omega \)--deformation of \( \mathcal{N} = 4 \) super-Yang--Mills.}
  \label{tab:D3-embedding-two-epsilon}
\end{table}

\printbibliography

\end{document}